\documentclass[fleqn,usenatbib]{mnras}

\usepackage{times,txfonts}

\usepackage[T1]{fontenc}
\usepackage{ae,aecompl}
\usepackage{xcolor}

\usepackage{graphicx}	
\usepackage{amsmath}	
\usepackage{amssymb}	

\title[The mass-radius relation]{The white dwarf mass--radius relation and its dependence on the hydrogen envelope.}

\author[Romero et al.]{Alejandra D. Romero,$^{1}$\thanks{E-mail: alejandra.romero@ufrgs.br}, S. O. Kepler$^{1}$, S. R. G. Joyce$^{2}$, G. R. Lauffer$^{1}$ 
\newauthor  \& A. H. C\'orsico$^{3,4}$
\\
$^{1}$Physics Institute, Universidade Federal do Rio Grande do Sul, Av. Bento Gon\c{c}alves 9500, Brazil\\
$^{2}$Dept. of Physics \& Astronomy, University of Leicester, University Road, Leicester, LE1 7RH\\
$^{3}$Facultad de Ciencias Astron\'omicas y Geof\'isicas, Universidad Nacional de La Plata, La Plata 1900, Argentina\\
$^{4}$CONICET, Consejo Nacional de Investigaciones Cientif\'icas y T\'ecnicas, Argentina
}

\date{Accepted XXX. Received YYY; in original form ZZZ}

\pubyear{2015}

\begin{document}
\label{firstpage}
\pagerange{\pageref{firstpage}--\pageref{lastpage}}
\maketitle

\begin{abstract}

We present a study of the dependence of the mass--radius relation for DA white dwarf stars on the hydrogen envelope mass and the impact on the value of $\log g$, and thus the determination of the stellar mass. We  employ a set of full evolutionary carbon-oxygen core white dwarf sequences with white dwarf mass between 0.493 and $1.05 M_{\odot}$. 
Computations of the pre-white dwarf evolution uncovers an intrinsic dependence of the maximum mass of the hydrogen envelope with stellar mass, i.e., it decreases when the total mass increases. 
We find that a reduction of the hydrogen envelope mass can lead to a reduction in the radius of the model of up to $\sim 12\%$. This translates directly into an increase in $\log g$ for a fixed stellar mass, that can reach up to 0.11 dex, mainly overestimating the determinations of stellar mass from atmospheric parameters. 
Finally, we find a good agreement between the results from the theoretical mass--radius relation and observations from white dwarfs in binary systems. In particular, we find a thin hydrogen mass of $M_H \sim 2\times 10^{-8} M_{\odot}$, for 40 Eridani B, in agreement with previous determinations. For Sirius B, the spectroscopic mass is 4.3\% lower than the dynamical mass. However, the values of mass and radius from gravitational redshift observations are compatible with the theoretical mass--radius relation for a thick hydrogen envelope of $M_H = 2 \times 10^{-6} M_{\odot}$.

\end{abstract}




\section{Introduction}

One of the results of Chandrasekhar's  theory for the structure of white
dwarf stars is a dependence of the radius with the stellar mass, known
as  the mass--radius  relation. 
This  relation is widely used in stellar astrophysics. It makes possible to estimate the stellar mass of white dwarf stars from   spectroscopic  temperatures   and  gravities, which in turn are used to determine the mass distribution \citep[see e.g.,][]{1979A&A....76..262K,                   2001ApJS..133..413B, 2005ApJ...630L..69L, 2012AJ....143...68H, 2012ApJ...757..116F, 2017MNRAS.465.2849T}. In addition, a determination of the white dwarf mass distribution contains information about the star formation history and is directly related with the initial--final mass function \citep{2008MNRAS.387.1693C,2016ApJ...818...84C, 2018ApJ...860L..17E} which determines how much stellar material is returned to the interstellar medium, affecting the chemical evolution of the Galaxy.  

Semi-empirical  determinations of  the mass--radius  relations can  be
obtained  from  atmospheric   parameters,  effective  temperature  and
surface gravity,  which combined with flux  measurements and parallax,
can  lead  to   a  determination  of  the  radius   and  stellar  mass
\citep{1998ApJ...494..759P,  2012AJ....143...68H,2017ApJ...848...11B}.
This technique started with the works by \citet{1996A&A...311..852S} and \citet{1997A&A...325.1055V}, who used atmospheric parameters and trigonometric parallax measurements for 20 white dwarfs observed with the  {\it Hypparcos} satellite. Later, this technique was expanded to include wide binary systems for which the primary has a precise parallax from  {\it Hipparcos} \citep{1998ApJ...494..759P,2012AJ....143...68H} and {\it Gaia} DR1 \citep{2017MNRAS.465.2849T}. However, this method is not completely independent of theoretical models, since the determination of the radius depends on the flux emitted at the surface of the star, that is based on the predictions of model atmospheres. In addition, the determinations of the effective temperature and surface gravity also rely on model atmospheres, usually through spectral fitting, which can suffer from large uncertainties, up to $\sim 0.1$ in $\log g$ and 1--10\% in temperature \citep{2018MNRAS.tmp.1366J,2019MNRAS.482.5222T}.


For eclipsing binary systems, the mass and radius of the white dwarf component can be obtained from photometric observations of the eclipses and kinematic parameters, without relying on white dwarf model atmospheres, except for the determination of the effective temperature. However, the specific configuration of eclipsing binaries implies that they have most probably interacted in the past, as common-envelope binaries \citep{2017MNRAS.465.2849T}. A sample of eclipsing binaries containing a white dwarf component applied to the study of the mass--radius relation can be found in \citet{2010MNRAS.402.2591P, 2012MNRAS.426.1950P, 2012MNRAS.420.3281P, 2017MNRAS.470.4473P}. In particular, \citet{2017MNRAS.470.4473P} analysed a sample of 16 white dwarfs in detached eclipsing binary and estimated their mass and radius up to a precision of 1--2 per cent. 

Another method to test the mass--radius relation is to rely on astrometric
binaries  with  precise orbital parameters, in particular a dynamical mass determination, and distances. Examples of those systems are Sirius, 40 Eridani and Procyon, for which recent determinations of the dynamical masses based on detailed orbital parameters were reported by \citet{2017ApJ...840...70B}, \citet{2017AJ....154..200M} and \citet{2015ApJ...813..106B}, respectively. However, the radius of the white dwarf component cannot be determined from orbital parameters and other techniques are necessary to estimate this parameters.  
In particular, for the systems mentioned above, the radius is estimated from the measured flux and precise parallax, depending on model atmospheres.


From evolutionary model computations for single stars it  is known that the theoretical 
mass--radius relation  depends systematically on effective temperature,
core composition, helium abundance and hydrogen abundance in the case of DA  white dwarf stars \citep{1995LNP...443...41W,2001PASP..113..409F,2010ApJ...717..183R,2010ApJ...716.1241S,2015MNRAS.450.3708R}. Previous theoretical mass--radius relations \citep[e.g.][]{1995LNP...443...41W,2001PASP..113..409F} have assumed a constant hydrogen layer thickness which is applied to all models regardless of progenitor and white dwarf mass, being typically   $M_H/M_*   =  10^{-4}$   \citep{1984ApJ...282..615I}. However, full evolutionary     computations     from     \citet{2012MNRAS.420.1462R,
  2013ApJ...779...58R} showed that the upper limit for the mass of the
hydrogen layer  in DA  white dwarf  depends on the  total mass  of the
remnant. The hydrogen  content can vary from $M_H/M_*  \sim 10^{-3}$, for
white dwarf  masses of $\sim 0.5  M_{\odot}$, to $M_H/M_* =  10^{-6}$ for
massive white dwarfs with $\sim 1 M_{\odot}$.
In    addition,
asteroseismological studies show strong evidence of the existence of a
hydrogen  layer  mass range  in  DA  white  dwarfs, within  the  range
$10^{-9.5}  < M_H/M_*  < 10^{-4}$,  with an  average of  $M_H/M_* \sim 10^{-6.3}$  \citep{2008PASP..120.1043F,2009MNRAS.396.1709C,
  2012MNRAS.420.1462R}. The mass of the hydrogen layer is an important factor, since the mass--radius relation varies by 1-15 per cent, depending on the white dwarf mass and temperature, whether a thick ($10^{-4}M_*$) or a thin ($10^{-10}M_*$) hydrogen layer is assumed \citep{2017MNRAS.465.2849T}.

In this  work we study the dependence of the mass--radius relation with the mass of the hydrogen layer.  The white  dwarf cooling sequences employed  are  those from  \citet{2012MNRAS.420.1462R, 2013ApJ...779...58R,
  2017ApJ...851...60R},   extracted   from   the   full   evolutionary
computations    using    the    {\tt   LPCODE}    evolutionary    code
\citep{2005A&A...435..631A,2010ApJ...717..183R}. The model  grid expands from $\sim 0.493
M_{\odot}$ to  $1.05 M_{\odot}$  in white dwarf mass,  where carbon-oxygen
core white dwarfs  are found. We also consider a range in hydrogen envelope mass from $\sim 10^{-3} M_{*}$  to $\sim  5\times 10^{-10}  M_{*}$, depending  on the  stellar mass.
 
We compare our theoretical sequences with mass and radius determinations for white dwarfs in binary systems, in order to test the predictions of the theoretical mass--radius relation and to measure the hydrogen content in the star, when possible. We consider four white dwarfs in astromeric binaries -- 40 Eridani B \citep{2017AJ....154..200M}, Sirius B \citep{2017ApJ...840...70B}, Procyon B \citep{2015ApJ...813..106B} and Stein 2051 B \citep{2017Sci...356.1046S} -- and a sample of 11 white dwarfs in detached eclipsing binaries \citep{2017MNRAS.470.4473P}. 

The paper is organized as follows. In  section  \ref{sequences}  we  briefly  describe  the  evolutionary
cooling  sequences  used in  our  analysis. Section \ref{burning} is devoted to study the evolution of the hydrogen mass in the white dwarf cooling sequence for different stellar masses. In section \ref{radio-H} we present an analysis on the dependence on the total radius of the white dwarf with the hydrogen envelope mass and the possible impacts on the spectroscopic stellar mass determinations. We also compare our theoretical cooling sequences with other model grids used in the literature. Section \ref{H-mass} is devoted to present the comparison between our theoretical models and the mass and radius obtained for white dwarfs in binary systems. Final  remarks  are  presented  in  section
\ref{end}.

\section{Computational details}\label{sequences}

\subsection{Input Physics}

The white dwarf cooling sequences employed in this work are those from
\citet{2012MNRAS.420.1462R, 2013ApJ...779...58R, 2017ApJ...851...60R},
extracted  from full  evolutionary  computations calculated with the  {\tt
  LPCODE}  evolutionary  code. Details on the code can be found in 
  \citet{2005A&A...435..631A, 2010ApJ...717..897A, 2010ApJ...717..183R, 2015MNRAS.450.3708R}.  
{\tt LPCODE} computes the evolution of single stars with low and intermediate mass at the main
sequence, starting at the zero age main sequence, going through the hydrogen and helium burning
stages, the thermally pulsing and mass-loss stages on the AGB, to the white dwarf cooling evolution.
Here we briefly mention the main input physics relevant for this work. 

The {\tt LPCODE} evolutionary  code considers a simultaneous treatment
of   non-instantaneous   mixing    and    burning     of    elements
\citep{2003A&A...404..593A}.  The  nuclear network accounts explicitly
for 16  elements and  34 nuclear reactions,  that include  $pp$ chain,
CNO-cycle,      helium      burning      and      carbon      ignition
\citep{2010ApJ...717..183R}.     

We  consider  the  occurrence  of extra-mixing beyond  each
convective     boundary      following     the     prescription     of
\citet{1997A&A...324L..81H},  except for  the thermally  pulsating AGB
phase. We treated the extra--mixing as a time--dependent diffusion process, assuming that the mixing velocities decay exponentially beyond each convective boundary. The diffusion coefficient is given by $D_{\rm EM} = D_0 \exp (−2z/f H_P )$, where $H_P$ is the pressure scaleheight at the convective boundary, $D_0$ is the diffusion coefficient of unstable regions close to the convective boundary, $z$ is the geometric distance from the edge of the convective boundary, and $f$ describes the efficiency, and was set to $f = 0.016$ \citep[see][for details]{2015MNRAS.450.3708R}.
Mass loss episodes follow the prescription from \citet{2005ApJ...630L..73S} during the core helium burning and the red  giant branch  phases,  and the prescription of \citet{1993ApJ...413..641V} during the AGB and  thermally pulsating AGB phases \citep{2017A&A...599A..21D,2018A&A...613A..46D}.
During  the white dwarf evolution, we
considered  the distinct  physical  processes that  modify the inner chemical
profile. In particular, element  diffusion strongly
affects  the   chemical  composition  profile   throughout  the  outer
layers. Indeed, our sequences  develop a pure hydrogen envelope with
increasing  thickness as  evolution  proceeds. Our  treatment of  time
dependent  diffusion  is based  on  the  multicomponent gas  treatment
presented  in \citet{1969fecg.book.....B}.  We  consider gravitational
settling  and thermal  and chemical  diffusion of  H,  $^3$He, $^4$He,
$^{12}$C, $^{13}$C, $^{14}$N and $^{16}$O \citep{2003A&A...404..593A}.
To  account for convection process  in the interior of the star, we adopted  the mixing
length theory,  in its  ML2 flavor, with  the free  parameter $\alpha=
1.61$ \citep{1990ApJS...72..335T} during the evolution previous to the 
white dwarf cooling curve, and $\alpha =1$ during the white dwarf evolution.   
Last, we considered  the chemical
rehomogenization  of  the   inner  carbon-oxygen  profile  induced  by
Rayleigh-Taylor instabilities following \citet{1997ApJ...486..413S}.

For the white dwarf stage, the  input  physics of  the  code includes  the  equation  of state  of
\citet{1994ApJ...434..641S} for  the high density  regime complemented
with   an   updated   version    of   the   equation   of   state   of
\citet{1979A&A....72..134M} for the low density regime. Other physical
ingredients  considered in  {\tt LPCODE}  are the  radiative opacities
from the OPAL opacity project \citep{1996ApJ...464..943I} supplemented
at    low    temperatures   with    the    molecular   opacities    of
\citet{1994ApJ...437..879A}.   Conductive  opacities  are  those  from
\citet{2007ApJ...661.1094C}, and the neutrino emission rates are taken
from \citet{1996ApJS..102..411I} and \citet{1994ApJ...425..222H}.

Cool white dwarf stars are expected to crystallize as a result of strong Coulomb interactions in their very dense interior \citep{1968ApJ...151..227V}. In the process two additional energy sources, i.e.,
the release of latent heat and the release of gravitational energy
associated with changes in the chemical composition of the
carbon--oxygen profile induced by crystallization \citep{1988Natur.333..642G,2009ApJ...693L...6W}, are considered self-consistently and locally coupled to the full set of equations of
stellar evolution. The chemical redistribution due to phase
separation has been considered following the procedure
described in \citet{1999ApJ...526..976M} and \citet{1997ApJ...486..413S}. To assess the enhancement of oxygen in the crystallized core we used the azeotropic-type formulation of \citet{2010PhRvL.104w1101H}.

\subsection{Model Grid}
\label{model-grid}

The DA white dwarf cooling sequences considered in this work are the result
of full evolutionary computations of progenitor stars with stellar masses between
$0.95$ and $6.6 M_{\odot}$ at the zero age main sequence. The initial metallicity 
was set to $Z=0.01$.  As a result, the stellar mass range in the cooling sequence 
expands  from $\sim  0.493  M_{\odot}$  to $1.05  M_{\odot}$, 
where carbon-oxygen  core white  dwarfs are  found. 
These sequences were presented in the works of \citet{2010ApJ...717..183R,2012MNRAS.420.1462R, 2013ApJ...779...58R, 2017ApJ...851...60R}.
The values of stellar mass of our model grid are listed in table \ref{masses}, along with the hydrogen and helium content as predicted by single stellar evolution,  for an effective temperature of $\sim 12\, 000$ K.
The central abundance
of carbon and oxygen for each mass is also listed. Note that the value of the hydrogen content 
listed in table \ref{masses} is the maximum possible value, since a larger hydrogen mass will 
trigger nuclear reactions, consuming all the exceeding material (see section \ref{burning}). The upper limit for the possible hydrogen content shows a strong dependence on the stellar mass. 
It ranges from $1.6\times 10^{-4} M_{\odot}$ for $M_* = 0.493 M_{\odot}$
to $1.5\times 10^{-6} M_{\odot}$ for $M_* = 1.050 M_{\odot}$, with a value $\sim 10^{-4}M_*$ for the averaged-mass
sequence of $M_* = 0.60 M_{\odot}$, for effective temperatures near the beginning of the ZZ Ceti instability strip. The helium abundance also shows a dependence with the stellar mass, decreasing monotonically with the increase of the stellar mass. In particular, the sequence with $1.05 M_{\odot}$ was obtained by artificially scaling the stellar mass from the $0.976 M_{\odot}$ sequence at high effective temperatures \citep[see][for details]{2013ApJ...779...58R}. Since no residual helium burning is present in the cooling sequence the helium content does not change, and both sequences present similar helium content. 

Uncertainties related to the physical processes occurring during the AGB stage, lead to uncertainties in the amount of hydrogen remaining on the envelope of a white dwarf star. For instance, the hydrogen mass can be reduced to a factor of two as a result of the carbon enrichment of the envelope due to third dredge--up episodes at the thermally pulsing AGB phase \citep{2015A&A...576A...9A}. Also, the hydrogen envelope mass depends on the initial metallicity of the progenitor, being  a factor of 2 thicker when the metallicity decreasses from $Z=0.01$ to $Z=0.001$ \citep{2010ApJ...717..183R,2015MNRAS.450.3708R}. However, the mass loss rate during the AGB and planetary nebula stages will not strongly impact the amount of hydrogen left on the white dwarf \citep{2015A&A...576A...9A}.
 
\begin{figure}
\includegraphics[width=\columnwidth]{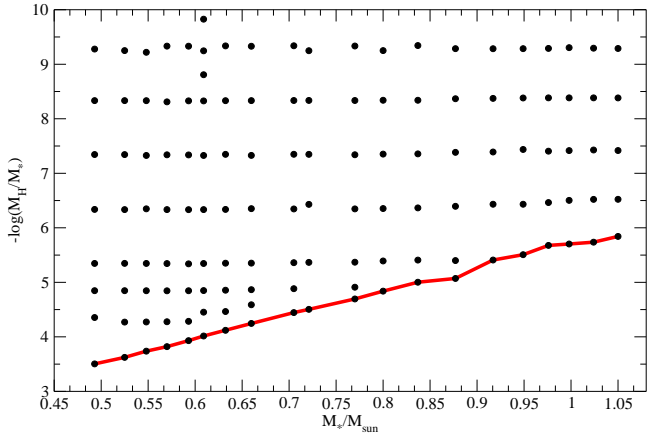}
\caption{Grid of DA WD evolutionary  sequences considered in this work
  in  the  $M_*/M_{\odot}$  vs. $-\log(M_H/M_*)$  plane.  Each  circle
  corresponds  to a  sequence  of models  representative  of white dwarf  stars
  characterized  by  a  given   stellar  mass  and  hydrogen  envelope
  mass. The envelope mass is measured at an effective temperature of $12\, 000$ K. The  red line connects  the sequences with the  maximum values
  for  the  thickness  of  the hydrogen  envelope,  predicted  by  our
  evolutionary computations. \label{massa-env}}
\end{figure}

 In order to compute cooling sequences with different values of the thickness of the hydrogen envelope, in particular thinner than the value expected by the burning limit,
 $^1$H was replaced with $^4$He at the bottom of the hydrogen envelope 
 \citep[see][for details]{2012MNRAS.420.1462R, 2013ApJ...779...58R}. This procedure is done at high effective 
 temperatures ($\gtrsim 90\, 000$K), so the transitory effects caused by the artificial procedure are quickly washed out.
The values  of hydrogen  content as  a function  of the  stellar mass  are
depicted in  Figure \ref{massa-env}. The  thick red line  connects the
values of the maximum value of $M_H$ predicted by our stellar evolution computations.



\begin{table}
\caption{The   main   characteristics   of    our   set   of   DA   white dwarf models. The stellar mass at the white dwarf stage is listed in column 1. Also listed are the hydrogen mass at $12\, 000$ K (column 2), the helium mass (column 3) and the central abundances of carbon (column 4) and oxygen (column 5).  }  \centering
\begin{tabular}{ccccc}
\hline\hline  $M_{\star}/M_{\sun}$  & $-\log(M_{\rm  H}/M_{\star})$  & $-\log(M_{\rm  He}/M_{\star})$ &  $X_{\rm C}$  & $X_{\rm  O}$\\
\hline
0.493 & 3.50 & 1.08 & 0.268 & 0.720\\ 
0.525 & 3.62 & 1.31 & 0.278  & 0.709\\ 
0.548 & 3.74 &  1.38 & 0.290 &  0.697\\  
0.570 & 3.82 & 1.46  & 0.301 & 0.696\\  
0.593 & 3.93 &  1.62 &  0.283 &  0.704\\ 
0.609 & 4.02 &  1.61 & 0.264  &  0.723\\  
0.632  &   4.25  &  1.76  &  0.234  & 0.755\\  
0.660 &  4.26 &  1.92  & 0.258  & 0.730\\  
0.705 &  4.45 &  2.12 &  0.326  &  0.661\\  
0.721  &   4.50  &  2.14  &  0.328  & 0.659\\  
0.770  & 4.70 & 2.23  & 0.332 & 0.655\\  
0.800 & 4.84 &  2.33 &  0.339 &  0.648\\ 
0.837 &  5.00 &  2.50 & 0.347  &  0.640\\  
0.878  &   5.07  &  2.59  &  0.367  & 0.611\\ 
0.917  & 5.41 & 2.88  & 0.378 & 0.609\\  
0.949 & 5.51 &  2.92 &  0.373 &  0.614\\ 
0.976 &  5.68 &  2.96 & 0.374  &  0.613\\  
0.998  &   5.70  &  3.11  &  0.358  & 0.629\\ 
1.024  & 5.74 & 3.25  & 0.356 & 0.631\\  
1.050 &  5.84 & 2.96 & 0.374 & 0.613\\ 
\hline\hline
\label{masses}
\end{tabular}
\end{table}

\section{The evolution of the hydrogen content}
\label{burning}

\begin{figure}
\includegraphics[width=\columnwidth]{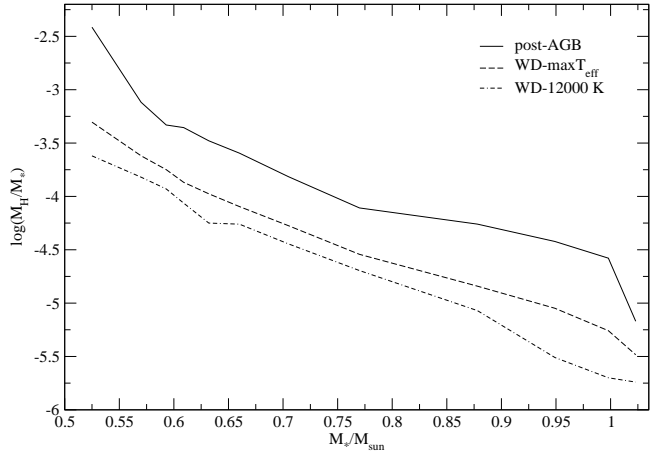}
\caption{Hydrogen mass as a function of the white dwarf stellar mass for three point during the evolution: during the nearly constant luminosity stage at the post-AGB at $T_{\rm eff} \sim 10\, 000$ K (post-AGB), at the point of maximum effective temperature when the star enters the cooling sequence (WD-maxT$_{\rm eff}$) and on the cooling curve at $T_{\rm eff}\sim 12\, 000$K (WD-12000K).
  evolutionary computations. \label{H-evol}}
\end{figure}

After the end of the TP-AGB stage, during the post-AGB evolution at nearly constant luminosity, simple models of the white dwarf progenitors show that CNO cycle reactions reduce the hydrogen content  below a critical value. If the star has a white dwarf mass of 0.6 $M_{\odot}$, the value  of the critical hydrogen mass is  $\sim 2\times 10^{-4} M_{\odot}$ \citep{1982ApJ...260..821I,1984ApJ...277..333I,1983ARA&A..21..271I}.
Residual
nuclear burning will reduce the hydrogen mass in the surface layers to
$\sim 10^{-4} M_{\odot}$. These values change with white dwarf mass when the evolution previous and during the cooling curve are computed consistently. In figure \ref{H-evol} we show the hydrogen mass as a function of the white dwarf mass for three points during the evolution: the point with $T_{\rm eff} \sim 10\, 000$ K during the post-AGB stage, previous to the white dwarf phase (solid line), the point of maximum effective temperature when the star enters the cooling curve (dashed line) and at $T_{\rm eff}\sim 12\, 000$K on the white dwarf cooling curve (dot-dashed line). As can be seen from figure \ref{H-evol}, the major reduction of the hydrogen mass occurs during the post-AGB evolution, for all stellar masses, due to CNO shell--burning. For a $M_{\rm WD} \sim 0.6 M_{\odot}$ the hydrogen content is $\sim 2.8\times 10^{-4}M_{\odot}$ at $10\, 000$ K on the post-AGB stage, and it reduces to  $\sim 1.1\times 10^{-4}M_{\odot}$ when the star enters the white dwarf cooling curve. 

\begin{figure}
\includegraphics[width=\columnwidth]{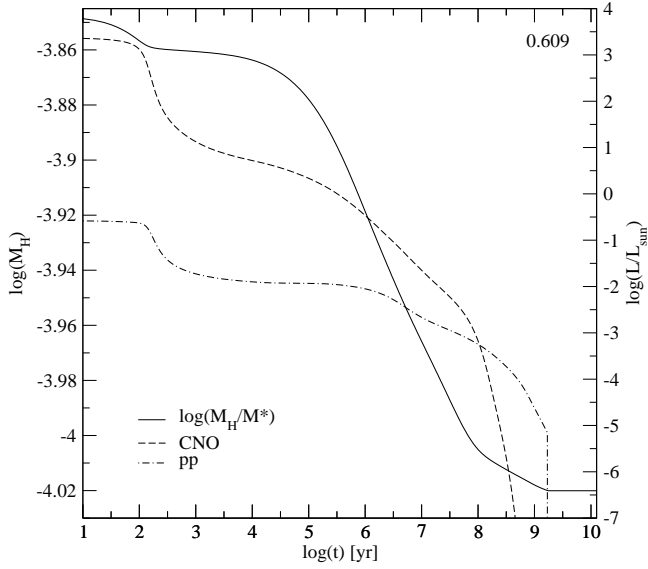}
\caption{Temporal evolution  of the  hydrogen content $M_H$  (in units of $M_*$)  and  the  ratio  of hydrogen  nuclear  burning  to  surface
  luminosity for  CNO bi-cycle and {\it pp} chain, for  a white
  dwarf  sequence  with  stellar  mass  0.609  $M_{\odot}$.  The  time
  corresponds to the cooling time in yr measured from the point of maximum effective temperature at the beginning of the cooling curve.\label{Mh-L-0609}}
\end{figure}

\begin{figure}
\includegraphics[width=\columnwidth]{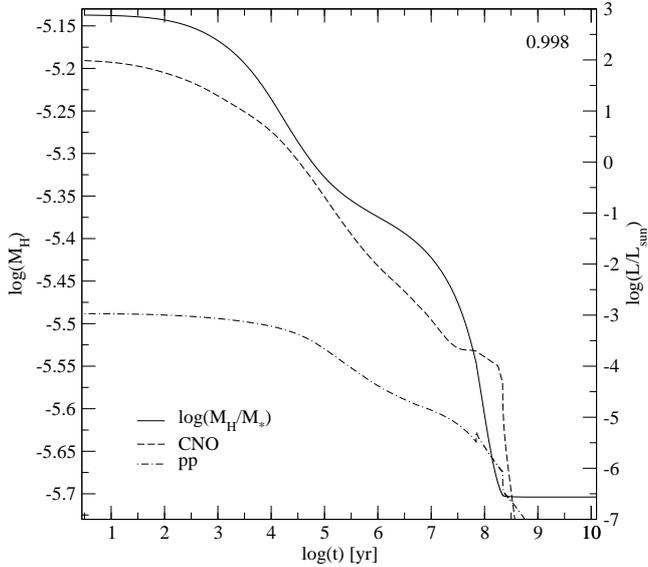}
\caption{Same  as  figure  \ref{Mh-L-0609}  but for  a  sequence  with
  stellar mass 0.998 $M_{\odot}$. The  time corresponds to the cooling
  time in yr.\label{Mh-L-0998}}
\end{figure}

Residual burning at the cooling curve can further reduce the hydrogen envelope  by a factor of $\sim$ 2 (see figure \ref{H-evol}). 
Figures   \ref{Mh-L-0609}  and   \ref{Mh-L-0998}  show   the  temporal
evolution  of the  hydrogen  content for sequences  with white dwarf
stellar mass 0.609 and 0.998 $M_{\odot}$, respectively, during the cooling curve. Also shown are
the  luminosity given  by the nuclear  burning of hydrogen due to the CNO
bi-cycle and  the {\it pp}  chain as a function of the logarithm of the cooling time in years, measured from the point of maximum effective temperature when the star enters the white dwarf cooling curve. As can  be seen, the hydrogen 
burning is reducing the hydrogen envelope mass during the early  evolutionary phases of the white dwarf stage. 
At  the  beginning  of the  cooling  sequence, the  0.609
$M_{\odot}$ sequence has a hydrogen content of $M_H/M_{\odot} = 8.6 \times 10^{-5}$, which is 
reduced due to residual hydrogen burning to a constant value of $M_H/M_{\odot} = 5.8\times 10^{-5}$
after 1.76 Gyr. As can be seen from figure \ref{Mh-L-0609}, that the CNO bi-cycle dominates de energy
production due to hydrogen burning for the first $\sim$110 Myr of the cooling sequences. Hydrogen-burning due to the {\it pp} chain lasts longer, causing a small reduction in the 
hydrogen content of the model. Once the hydrogen content decreases below a certain threshold
the pressure at the bottom of the envelope is not large enough to support further nuclear reactions.
After no residual nuclear burning is present in the star, the main energy source of the white dwarf is 
the release of gravothermal energy.
A similar scenario is found for the more massive sequence shown in figure \ref{Mh-L-0998}. 
The star enters the cooling sequence with an hydrogen content of $M_H/M_{\odot} = 7.3 \times 10^{-6}$,
10 times smaller than the hydrogen content in the 0.609 $M_{\odot}$ model at the same stage. After 580 Myrs 
the hydrogen content reaches a somewhat constant value of $M_H/M_{\odot} = 1.97\times 10^{-6}$.
In this case, because of the higher temperature at the base of the envelope, the contribution to the 
energy production from the CNO bi-cycle is five orders of magnitude larger than the contribution from
the proton-proton chain at the beginning of the cooling sequence, being the dominant source while residual 
nuclear reactions are still active. 

\subsection{Massive white dwarf with "thick" hydrogen envelope} 
\label{acretion-1}

\begin{figure}
\includegraphics[width=\columnwidth, angle=0]{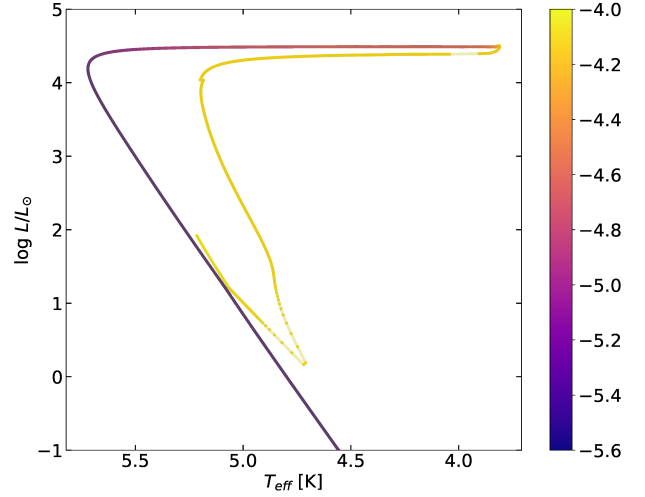}
\caption{Evolution in the H-R diagram for a sequence of 1$M_{\odot}$ in the white dwarf cooling sequence with an artificially increased hydrogen envelope of $M_{\rm H} \sim 10^ {-4} M_*$ at the begining of the cooling curve. This scenario sumilates the possible acretion of material due to interaction with a binary companion. The color bar indicates the amount of hydrogen left in the envelope of the star. Note that, if residual thermonuclear burning is considered, the amount of hydrogen decreases and reaches a value of  $\sim 2.5 \times 10^{-6} M_*$, similar to that obtained from single stellar evolution.
\label{acretion}}
\end{figure}

The hydrogen content in the envelope of a white dwarf can increase due to accretion from a companion star.
Depending on the distance to the companion, a certain amount of mass can be added on top of the white dwarf.
However, if the hydrogen content exceeds the limiting value for nuclear reactions, the H--shell at the bottom of the hydrogen
envelope can be activated and the additional hydrogen is consumed, leading to an equilibrium mass. 
To explore this scenario, we computed the cooling evolution of a 1$M_{\odot}$ white 
dwarf sequence with a thick hydrogen layer. To simulate the accreted material we artificially increased the hydrogen 
envelope mass at high effective temperatures, near the beginning of the cooling sequence. For a white dwarf with 1$M_{\odot}$ the remaining hydrogen content
predicted by single stellar evolution is $M_H/M_* = 2.1 \times 10^{-6}$ (see Table. \ref{masses}). Thus, we increased the 
hydrogen mass to a factor 100, $M_H/M_* = 10^{-4}$, and computed the following evolution considering possible sources of nuclear 
burning. Figure \ref{acretion} shows the evolution in the H--R diagram for this sequence. The color bar indicates the hydrogen envelope mass in a logarithmic scale. As expected, the hydrogen burning shell at the bottom of the hydrogen envelope is active again, consuming the excess in the 
hydrogen content. The residual burning makes the model move to the high luminosity and low temperature region of the H--R 
diagram, similar to what happens in the low mass regime that produces pre-ELM white dwarf stars \citep{2013A&A...557A..19A,2014A&A...571A..45I}. Once the hydrogen content is reduced below the limiting value for nuclear 
burning, the star settles onto the cooling sequence one more time with a hydrogen mass of $\sim 2.5 \times 10^{-6} M_*$, similar to the 
value obtained from single evolution computations. Therefore, the final hydrogen content would not change significantly due to 
accretion.

\section{Radius and hydrogen content}
\label{radio-H}

\begin{figure}
\includegraphics[width=\columnwidth]{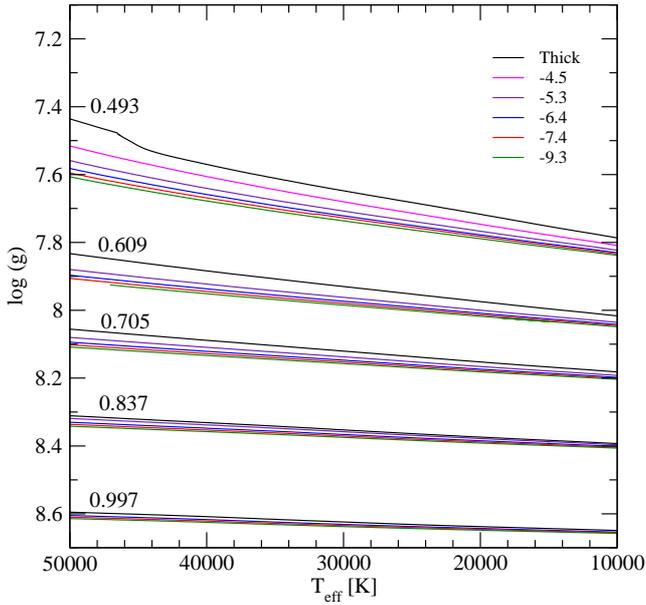}
\caption{Cooling  tracks  for  different  hydrogen  envelopes  in  the
  $T_{\rm eff}-\log g$  plane. The thickness of  the hydrogen envelope
  decreases from top  to bottom, and it is labelled  in the figure in $\log(M_{\rm H}/M_*)$. The
  stellar     mass    for     each    group     of    sequences     is
  indicated. \label{tracks1}}
\end{figure}

As it was mentioned, the mass-radius relation depends on the amount of hydrogen 
and helium present in the star. In particular, the hydrogen envelope, although it contains a very small amount of mass, appears to be the dominant parameter in the determination of the radius for DA white dwarfs \citep{2017MNRAS.465.2849T}. 
Thus a reduction of the hydrogen, and/or helium content, will lead to a smaller radius and thus to an 
increase in the surface gravity. This effect is depicted in figure \ref{tracks1}, where we show white dwarf evolutionary
sequences in the $\log g -T_{\rm eff}$ plane for stellar masses from 0.493 to 0.998$M_{\odot}$. The black line
indicates the sequence with the thickest hydrogen envelope, as predicted by single stellar evolution. As expected, 
the surface gravity increases when the hydrogen content decreases for a given stellar mass at a given temperature.
The effect is less important for higher stellar masses, since they form with thinner hydrogen envelopes. For instance, the maximum 
mass for the hydrogen content in a 0.998 $M_{\odot}$ sequence is 100 times thinner than that corresponding to a 0.6 $M_{\odot}$
(see table. \ref{masses} for details). The radius and $\log g$ for the stellar masses are listed in Table \ref{radius}
for effective temperatures of $40\, 000$ and $20\, 000$ K.

Comparing the results for the sequences with the thickest envelope 
with those having the thinnest hydrogen envelope of the grid, the reduction in the total radius is between 8-12\% for a stellar 
mass of 0.493 $M_{\odot}$, 5-8\% for stellar mass 0.609 $M_{\odot}$ and almost negligible, 1-2\%, for a model with 
0.998 $M_{\odot}$, within the ranges of hydrogen mass considered in this work. However, this reduction in the total radius has a strong impact in the surface gravity value, specially for 
high effective temperatures, being 0.11 dex, 0.08 dex and 0.02 dex in $\log g$ for stellar masses 0.493, 0.609 and 0.998 $M_{\odot}$, 
respectively, for $T_{\rm eff} = 40\, 000$ K. For an effective temperature of $20\, 000$ K, de increase in $\log g$ is 0.08, 0.07 and 
0.011 dex, respectively. Considering that the real mean uncertainty from spectroscopic fits in $\log g$ is $\sim 0.038$ dex 
\citep{2005ApJ...630L..69L,2005MNRAS.362.1134B}, it is possible to estimate the hydrogen layer for stellar masses lower than $\sim 0.7 M_{\odot}$, but not for higher stellar masses, unless the uncertainties are reduced to less than 0.011 dex in $\log g$.

\begin{table}
\caption{Radius (in $R_{\odot}$) and surface gravity ($g$ in cm/s$^2$) extracted from theoretical cooling sequences, for $40\, 000$ K (columns 2 and 3),
and $20\, 000$ K (columns 4 and 5), for the stellar masses presented in Figure \ref{tracks1}. }  \centering
\begin{tabular}{ccccc}
\hline\hline 
0.493  & $R(40 kK)$  & $\log g$  & $R(20 kK)$  & $\log g$\\
\hline
thick & 0.019137 & 7.5683 & 0.016157 & 7.7153\\
-5.3  & 0.017661 & 7.6380 & 0.015266 & 7.7646\\
-6.4  & 0.017222 & 7.6586 & 0.014986 & 7.7793 \\
-7.4  & 0.017017 & 7.6690 & 0.014927 & 7.7828 \\
-9.3  & 0.016826 & 7.6788 & 0.014814 & 7.7894 \\
\hline\hline
0.609 & $R(40 kK)$  & $\log g$  & $R(20 kK)$  & $\log g$\\
\hline
thick & 0.014794 & 7.8825 & 0.013306 & 7.9747\\
-5.3  & 0.014181 & 7.9206 & 0.012982 & 7.9973\\
-6.4  & 0.013899 & 7.9368 & 0.012796 & 8.0086\\
-7.4  & 0.013766 & 7.9451 & 0.012723 & 8.0136\\
-9.3  & 0.013658 & 7.9520 & 0.012662 & 8.0177\\
\hline\hline
0.705 & $R(40 kK)$  & $\log g$  & $R(20 kK)$  & $\log g$\\
\hline
thick & 0.0125432 & 8.0891 & 0.0116590 & 8.1526\\
-5.3  & 0.0122555 & 8.1093 & 0.0114827 & 8.1658\\
-6.4  & 0.0120894 & 8.1211 & 0.0113798 & 8.1737\\
-7.4  & 0.0120037 & 8.1273 & 0.0113274 & 8.1777\\
-9.3  & 0.0119206 & 8.1334 & 0.0112778 & 8.1815\\
\hline\hline
0.837 & $R(40 kK)$  & $\log g$  & $R(20 kK)$  & $\log g$\\
\hline
thick & 0.0103397 & 8.3315 & 0.0098452 & 8.3744\\
-5.3  & 0.0110280 & 8.3365 & 0.0098002 & 8.3781\\
-6.4  & 0.0101528 & 8.3474 & 0.0097275 & 8.3846\\
-7.4  & 0.0100880 & 8.3530 & 0.0096864 & 8.3882\\
-9.3  & 0.0100311 & 8.3579 & 0.0096513 & 8.3914\\
\hline\hline
0.998 & $R(40 kK)$  & $\log g$  & $R(20 kK)$  & $\log g$\\
\hline
thick & 0.008221 & 8.6082 & 0.007945 & 8.6377\\
-6.4  & 0.008124 & 8.6172 & 0.007886 & 8.6430\\
-7.4  & 0.008084 & 8.6215 & 0.007860 & 8.6458\\
-9.3  & 0.008050 & 8.6251 & 0.007837 & 8.6484\\
\hline\hline
\label{radius}
\end{tabular}
\end{table}

\subsection{Comparison with other theoretical models}\label{comparison}

In the literature we can find a few grids of theoretical white dwarf cooling sequences. \citet{1995LNP...443...41W} computed DA white dwarf cooling sequences to study the white dwarf luminosity function of our Galaxy. He considered a stratified model with various central compositions, from pure C to pure O. The computations start as polytropes and the early evolution is characterized by a contraction phase at a constant luminosity of $\sim 10^2 L_{\odot}$, before entering the cooling curve \citep{1987ApJ...315L..77W}. The growing degeneracy in the core halts the contraction and the surface temperature reaches a maximum of $T_{\rm eff} \geq 100\, 000$ K. In particular, the models presented in \citet{1995LNP...443...41W} have a fixed hydrogen content of $\sim 10^{-4} M_*$.


Another widely used set of white dwarf sequences are those computed by the Montreal  group\footnote{http://www.astro.umontreal.ca/$\sim$bergeron/CoolingModels/}, and published in \citet{2001PASP..113..409F}. The first set of models for DA white dwarfs had a C pure core and a helium and hydrogen content of $10^{-2}M_*$ and $10^{-4}M_*$, respectively, for sequences with stellar mass in the range between 0.2 and 1.3 $M_{\odot}$. Latter, additional models with a C/O = 50/50 where computed with a helium content $10^{-2}M_*$ and two different values for the hydrogen mass, being $10^{-4}M_*$ ("thick") and $10^{-10}M_*$ ("thin").


Finally, \citet{2010ApJ...716.1241S} presented a set of white dwarf cooling sequences with stellar masses between 0.54 and 1 $M_{\odot}$. 
For each white dwarf mass an initial model was converged at $L\sim 10^2 L_{\odot}$, considering a chemical composition profile taken from pre-white dwarf computations, specifically at the first thermal pulse \citep{2000ApJ...544.1036S}. The hydrogen and helium mass are set to be $M_H =  10^{-4} M_*$ and $M_{He}=10^{-2}M_*$, respectively for all stellar masses, as in \cite{2001PASP..113..409F}.

In all cases, the authors assume a constant thickness of the hydrogen layer for all stellar masses. In order to keep the hydrogen mass at a strictly constant value, no residual thermonuclear burning has been included in the calculations. In particular, \citet{2010ApJ...716.1241S} stated that H--burning at the bottom of the hydrogen envelope is negligible in all but the more massive models. In addition, the cooling sequences presented in those works do not compute the post-AGB and planetary nebula stages, crucial to determine the hydrogen envelope mass at the white dwarf stage.

As we show in section \ref{model-grid}, the maximum mass of hydrogen left on top of a white dwarf model depends on the stellar mass. In particular,  the hydrogen envelope mass is larger than $10^{-4}M_*$ for stellar masses lower than $\sim 0.6 M_{\odot}$, while sequences with white dwarf masses larger than $\sim 0.6 M_{\odot}$ show thinner hydrogen envelopes, with masses below $10^{-4}M_*$. The extension of the hydrogen envelope will impact the total radius of the star and consequently the value of $\log g$. For instance, the hydrogen envelope mass for a $\sim 1 M_{\odot}$ white dwarf model is $\sim 10^{-6} M_*$, 100 times thinner than the fixed value considered in previous works, leading to a $\sim 5\%$ decrease in the stellar radius for that stellar masses and radii.  

\begin{figure}
\includegraphics[width=\columnwidth]{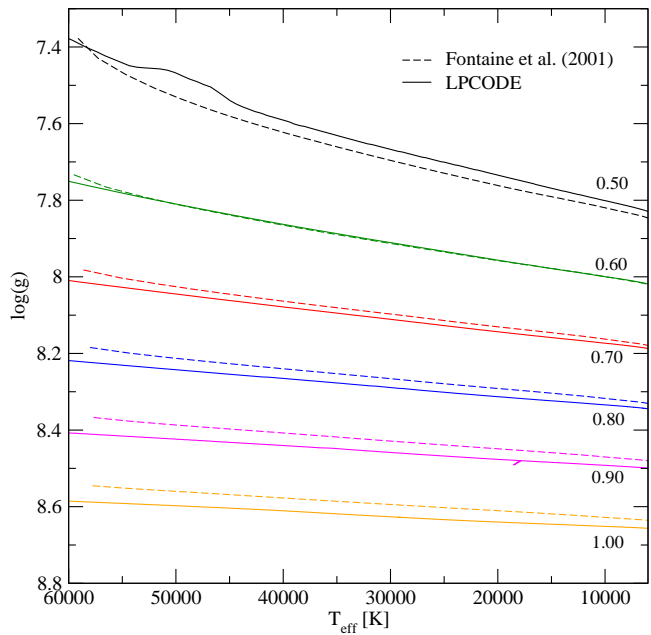}
\caption{Comparison in the $T_{\rm eff}-\log g$ plane of the theoretical cooling sequences computed with {\tt LPCODE} (solid lines) and 
those extracted from \citet{2001PASP..113..409F} with $M_H=10^{-4}M_*$ (dashed lines). The labels indicates the stellar mass of the sequences. In order to match  the stellar mass values from \citet{2001PASP..113..409F}, we interpolated the cooling sequences within our model grid (see Table \ref{masses}).  \label{tracks-comp}}
\end{figure}

In figure \ref{tracks-comp} we compare our canonical sequences, those with the thickest hydrogen envelope
obtained from single stellar evolution, to the theoretical cooling sequences from \citet{2001PASP..113..409F} with $M_H=10^{-4}M_*$, in the $T_{\rm eff}-\log g$ plane. Similar results are found when we compare to the cooling sequences from \citet{1995LNP...443...41W} and \citet{2010ApJ...716.1241S}.
To match the stellar mass values from \citet{2001PASP..113..409F}, we interpolated the cooling sequences within our model grid. 
Note that for a stellar mass 0.6$M_{\odot}$ the model computed with the {\tt LPCODE} perfectly overlaps 
with the model computed by \citet{2001PASP..113..409F}. This is a consequence of the value of the hydrogen envelope mass, which for this stellar mass is nearly $\sim 10^{-4} M_*$ in both cases. For stellar masses below 0.6 $M_{\odot}$, in this case 0.5
$M_{\odot}$, the cooling sequence computed with {\tt LPCODE} shows a lower $\log g$ -- larger radius -- than that from  \citet{2001PASP..113..409F}, 
while the opposite happens for sequences with stellar masses larger than  0.6 $M_{\odot}$. Thus, models with fixed hydrogen envelopes, that do not consider the pre-white dwarf evolution and/or the residual burning sources at the cooling sequence, can lead to overestimated or underestimated spectroscopic masses.

\section{Measuring the hydrogen mass}\label{H-mass}

In this section we estimate the hydrogen mass content of a selected sample of white dwarf stars in binary systems. 
The mass and radius for the objects considered in our analysis were taken from the literature and were estimated using different techniques, which are in principle, independent of the theoretical 
mass--radius relation. First we consider two well studied members of astrometric binary systems, 40 Eridani B and Sirius B. Aditionally, we consider two non-DA white dwarfs, Procyon B and Stein 2051 B. Finally, we analyse the results obtained from a sample of eleven detached eclipsing binaries. We do not consider the data fully based on spectroscopic techniques since the uncertainties in the atmospheric parameters, specially in $\log g$, is too large to estimate the hydrogen envelope mass \citep{2018MNRAS.tmp.1366J}. In each case we analyse the results and the uncertainties 
and how they impact the determination of the hydrogen layer mass.

\subsection{Astrometric binaries}

Astrometric binaries are an important tool to test the mass-radius 
relation, since independent determinations of the stellar mass can be obtained from the dynamical parameters of the 
binary system and accurate distances. The determination of the radius, on the other hand, is not model-independent, since it depends on the value of the flux emitted at the surface itself. We consider that, for these systems, the dynamical parameters and
distances are precise enough to set constrains not only on the theoretical mass--radius relation but on the hydrogen content.
In this section we use the observational determinations of mass and radius for four white dwarfs in astrometric binary systems.

\subsubsection{40 Eridani B}\label{40EriB}

40 Eridani B was for many years reported as a low mass white dwarf with a
stellar     mass    of     $\sim     0.4    M_{\odot}$.      Recently,
\citet{2017AJ....154..200M} determined  a dynamical mass  of $0.573\pm
0.018 M_{\odot}$  using observations of the orbit covering a longer period of time and the updated {\it  Hipparcos} parallax. Latter,
\citet{2017ApJ...848...16B} determined  the atmospheric  parameters of
this star using spectroscopy and  obtained an effective temperature of
$17\, 200  \pm 110$  K and  a surface  gravity of  $\log g  = 7.957\pm
0.020$. In addition,  these authors determined the radius of  40 Eridani B
using photometric  observations combined with the  distance \citep[see
  for   details][]{2017ApJ...848...11B}, being  $R=0.01308\pm 0.00020 R_{\odot}$.

\begin{figure}
\includegraphics[width=\columnwidth]{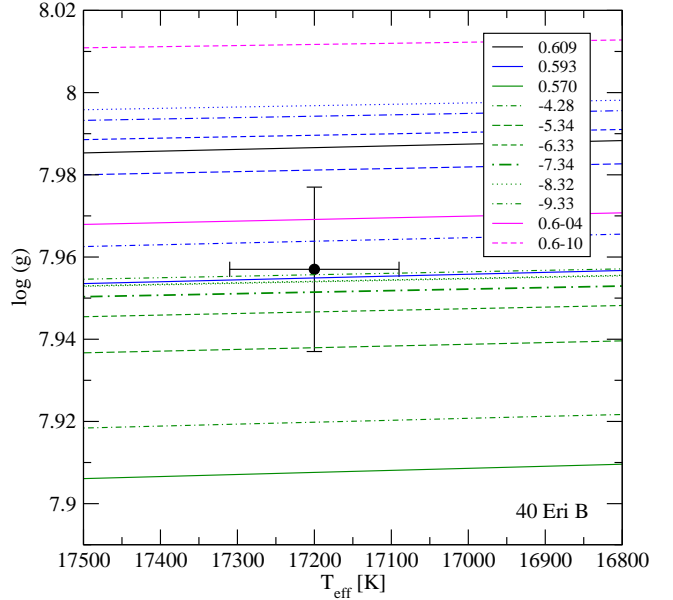}
\caption{Location of 40 Eridani B in  the $T_{\rm eff}-\log g$ plane. Solid
  lines correspond to thick  envelope sequences, those with the larger amount of hydrogen allowed by single stellar evolution. Thinner envelopes are depicted with different lines (see  inset in the figure). The labels
  associated to the sequences with different thin envelopes correspond
  to the value of $\log(M_H/M_*)$. Different colors indicate different
  stellar masses (in solar units).  Green and blue lines correspond to
  sequences with  0.570 and 0.593  $M_{\odot}$, respectively, while the  black solid 
  line correspond to  a sequences with $0.609  M_{\odot}$. The magenta
  lines corresponds to sequences from \citet{2001PASP..113..409F}
  for a  stellar mass  $0.6 M_{\odot}$ with  thick (0.6--04)  and thin
  (0.6--10) hydrogen envelopes. \label{40EriB-1}}
\end{figure}

From the spectroscopic parameters derived by \citet{2017ApJ...848...16B}, 
we computed the  stellar mass  for 40  Eridani B  from  
evolutionary tracks in  the $T_{\rm  eff}-\log g$ plane for different values of the hydrogen envelope thickness.  
In Figure \ref{40EriB-1} we depict the location
of 40 Eridani B along with theoretical white dwarf cooling tracks computed
with {\tt LPCODE}, with  different hydrogen envelope thickness ranging
from $10^{-4} M_*$ to $2\times 10^{-10} M_*$, and stellar mass between
0.570 and  0.609 $M_{\odot}$.  We also  plotted two  cooling sequences
from  \citet{2001PASP..113..409F}  with   $0.6  M_{\odot}$  for  thick
($10^{-4}M_*$) and thin ($10^{-10}M_*$) hydrogen envelopes.

If  we  consider  the  uncertainties in  the  atmospheric  parameters,
specifically in $\log  g$, the stellar mass varies  when the different
hydrogen     envelope    thickness     is    taken     into    account
\citep{2017MNRAS.465.2849T}. From  figure \ref{40EriB-1} we  note that
the spectroscopic stellar mass is higher if we consider thick envelope
tracks,  those with  the  thickest hydrogen  envelope  allowed by  our
models  of single  stellar  evolution, and  it  decreases for  thinner
hydrogen envelopes.  We computed  the stellar  mass for  each envelope
thickness. The results  are listed in Table  \ref{masa-40}, along with
the corresponding hydrogen  mass in solar units, the determinations of the dynamical mass \citep{1997ApJ...488L..43S,2017AJ....154..200M} and the spectroscopic mass \citep{2017ApJ...848...16B}. We notice that
the spectroscopic stellar mass varies  from $0.594 M_{\odot}$, for the
thick envelope set of tracks,  to $0.571 M_{\odot}$, $\sim 4$\% lower,
for the  thinnest value. Note that the values for the spectroscopic mass for the two thinnest envelopes are the same, implying that this parameter is not sensitive to the hydrogen envelope once it is thinner than $2.79\times 10^{-9} M_{\odot}$. The  spectroscopic mass that  better matches
the value  of the  dynamical mass from  \citet{2017AJ....154..200M} is
the one  characterized by an hydrogen  envelope of $M_H =  2.63 \times
10^{-8}M_{\odot}$.  If  we  consider  1  $\sigma$  uncertainties,  the
hydrogen mass is between $M_H = 2.63 \times 10^{-6}M_{\odot}$ and $M_H
= 2.67 \times 10^{-10}M_{\odot}$. Thus,  we conclude that the hydrogen
envelope for  40 Eridani B should  be thinner than the  value predicted by
single   stellar   evolution.  This result is consistent with previous works \citep{2012AJ....143...68H,2017ApJ...848...11B,2017ApJ...848...16B}. In particular, \citet{2017ApJ...848...16B}   found   a
spectroscopic mass  somewhat lower than  the values presented  in this
work, but  they also found  a thin envelope  for 40 Eridani  B, consistent
with $M_H=10^{-10}M_{\odot}$.

\begin{table*}
\renewcommand{\thetable}{\arabic{table}} \centering
\caption{Stellar mass  determinations for different  hydrogen envelope
  layers, considering the spectroscopic  atmospheric parameters for 40
  Eridani B  (left) and Sirius  B (right).  Also listed are  the dynamical
  mass \citep{2017AJ....154..200M} and  the spectroscopic mass derived
  by   \citet{2017ApJ...848...16B}.   Dynamical   masses   and   other
  spectroscopic   determinations   are   listed  in   the   last   three  rows.
  Note: Values of the stellar mass determined by the techniques: $^{(1)}$ dynamical mass, $^{(2)}$ fully spectroscopical, $^{(3)}$ gravitational redshift.} \label{masa-40}
\begin{tabular}{cc|cc}
\hline \hline $M_H/M_{\odot}$ & Mass [$M_{\odot}]$ & $M_H/M_{\odot}$ & Mass [$M_{\odot}]$\\ 
\hline 
$6.87\times 10^{-5}$ & $0.594 \pm 0.010$ & $\cdots$  & $\cdots$  \\ 
$3.07\times  10^{-5}$ &  $0.589 \pm  0.010$ & $\cdots$  & $\cdots$\\  
$2.64\times  10^{-6}$ &  $0.580  \pm 0.010$  & $2.02\times 10^{-6}$  & $0.974  \pm 0.013$  \\ 
$2.67\times  10^{-7}$ & $0.575  \pm  0.010$  &  $3.16\times   10^{-7}$  &  $0.970  \pm  0.015$ \\ 
$2.63\times 10^{-8}$  & $0.573 \pm 0.011$ &  $3.66\times 10^{-8}$ & $0.968  \pm 0.014$  \\  
$2.79\times  10^{-9}$ &  $0.571  \pm 0.011$  & $4.00\times  10^{-9}$ &  $0.967 \pm  0.014$\\ 
$2.65\times  10^{-10}$ & $0.571  \pm  0.011$  &  $4.87\times  10^{-10}$  &  $0.966  \pm  0.014$ \\   
\hline  
\citet{1997ApJ...488L..43S}$^{(1)}$ & $0.501 \pm 0.011$ & \citet{2017ApJ...840...70B}$^{(1)}$ & $1.018\pm 0.011$\\   
\citet{2017AJ....154..200M}$^{(1)}$ & $0.573 \pm 0.018$ & \citet{2005MNRAS.362.1134B}$^{(2)}$ & $0.978 \pm 0.005$\\ 
\citet{2017ApJ...848...16B}$^{(2)}$ & $0.565 \pm 0.031$ & \citet{2018MNRAS.tmp.2291J}$^{(3)}$ & $1.017\pm 0.025$\\
\hline
\end{tabular}
\end{table*}

\begin{figure}
\includegraphics[width=\columnwidth]{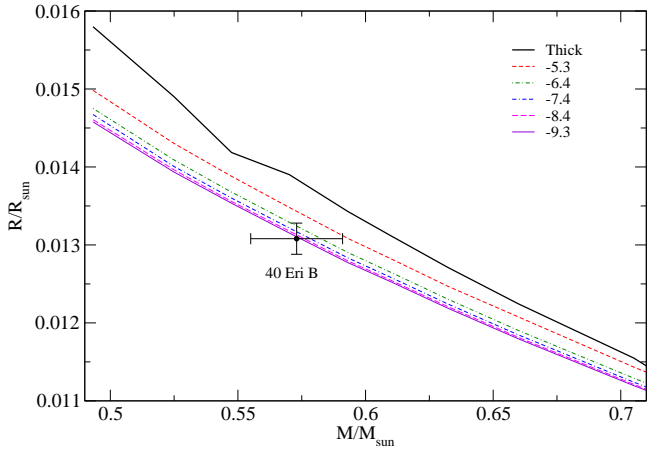}
\caption{Location of 40 Eridani B in  the mass--radius plane, where the values correspond to 
the dynamical mass \citep{2017AJ....154..200M} and radius \citep{2017ApJ...848...16B}. The curves are
  the  mass--radius relation  for an  effective temperature  of $T_{\rm
    eff}= 17\, 200$ K,  corresponding to the spectroscopic temperature
  of  40 Eridani  B.  Each curve  is  characterized by  a  value of  $M_H/M_*$.  The  black  solid  line ("Thick")  corresponds  to  the
 maximum hydrogen content allowed by single stellar evolution. \label{40EriB-2}}
\end{figure}

Another way to estimate the hydrogen mass in 40 Eridani B is by comparing
the  radius  and  dynamical   mass  to  the  theoretical  mass--radius
relation. This is shown in  figure \ref{40EriB-2}, where we depict the
mass--radius relations for six different envelope thickness. The solid
black curve correspond to the mass--radius relation for
thick  envelope sequences. The  thinnest
hydrogen  envelope  in the  model  grid  is $\log(M_H/M_*)\sim  -9.33$.  From figure  \ref{40EriB-2}, we see
that  the observations  of  40 Eridani  B  are in  agreement  with a  thin
envelope solution. Specifically, the upper limit for the hydrogen mass
is the same as the one obtained  using the value of the dynamical mass
and the spectroscopic parameters (Figure \ref{40EriB-1}).

The cooling age  for 40  Eridani B, for a stellar mass of $(0.573 \pm 0.0011) M_{\odot}$,   is $\sim  145$ Myrs. Considering the 
Initial--to--final mass relation  from \citet{2015MNRAS.450.3708R} for
solar  metallicity, we  estimate a  progenitor mass  of $1.53\pm  0.11
M_{\odot}$ and a total age of $2.68\pm 0.43$ Gyr for 40 Eridani B.

\subsubsection{Sirius B}\label{siriusB}

Sirius B is the brightest and  nearest of all white dwarfs, located at 2.65 pc.
Combining the information from the orbital parameters and
parallax, \citet{2017ApJ...840...70B} determined a dynamical mass of $M_{\rm B}
= 1.018 \pm  0.011 M_{\odot}$ for Sirius B. In addition,  \citet{2017ApJ...840...70B} 
reported spectroscopic atmospheric parameters  of $T_{\rm
  eff} = 25\, 369  \pm 46$ K and $\log g = 8.591  \pm 0.016$ and a  
radius of $R= 0.008098 \pm 0.000046 R_{\odot}$. Note that, the surface  gravity calculated from radius and
dynamical mass is $\log g =  8.629 \pm 0.007$, not compatible with the
spectroscopic value within 2 $\sigma$ \citep{2017ApJ...840...70B}. More recently, \cite{2018MNRAS.tmp.2291J} determine the mass for Sirius B using the effect of gravitational redshift and a radius from the flux and the parallax, and obtained values of $M_* = 1.017 \pm 0.025 M_{\odot}$  and $R_* = 0.00803\pm 0.00011 R_{\odot}$, in agreement with those of \citet{2017ApJ...840...70B}.

\begin{figure}
\includegraphics[width=\columnwidth]{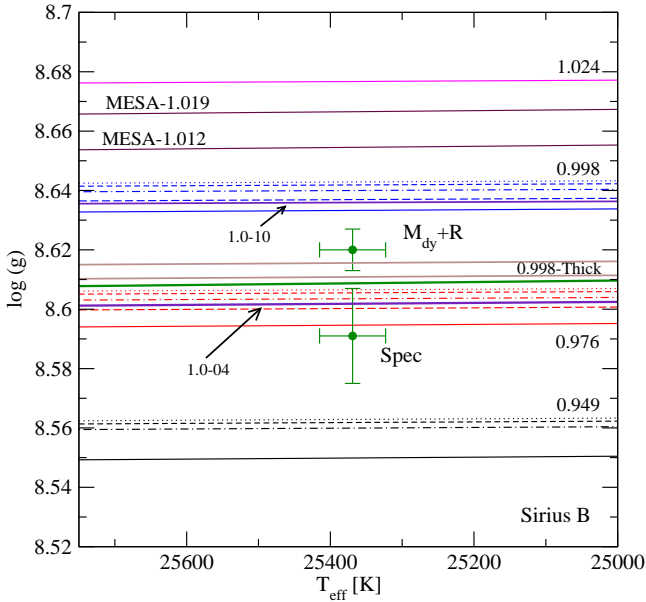}
\caption{Location of Sirius B in the $T_{\rm eff}-\log g$ plane, using
  the  spectroscopic  determinations  (Spec) and  the  dynamical  mass
  combined     with    the     radius     (M$_{\rm    dy}$+R)     from
  \citet{2017ApJ...840...70B}. The lines correspond to theoretical white dwarf sequences, that are color-coded in stellar mass: black
  lines for $0.949  M_{\odot}$, red lines for  $0.976 M_{\odot}$, blue
  lines   for  $0.998   M_{\odot}$  and   magenta  lines   for  $1.024
  M_{\odot}$. Solid  lines correspond to canonical  sequences. Thinner
  envelopes are  depicted with different  lines, being thinner  as the
  $\log  g$ increases.  The  violet lines  correspond  to the  cooling
  tracks from  \citet{2001PASP..113..409F} for a stellar  mass of $1.0
  M_{\odot}$   with  thick   (1.0-04)  and   thin  (1.0-10)   hydrogen
  envelopes. Just for comparison, we included two tracks 
  computed    with   the    MESA   code with stellar masses 1.012 and 1.019 $M_{\odot}$ from  \citet{2018MNRAS.480.1547L}. The thick  brown line labelled "0.998-Thick"
  correspond  to a cooling track computed  with  {\tt LPCODE}  having
  thicker hydrogen  envelopes than the  canonical value (see  text for
  details).\label{SiriusB-1}}.
\end{figure}

Using      the      spectroscopic     parameters      reported      in
\citet{2017ApJ...840...70B} we proceed to estimate the stellar mass of
Sirius B. In Figure \ref{SiriusB-1} we depict the location of Sirius B
in the $T_{\rm eff}-\log g$  plane.  Cooling tracks for stellar masses
in the range of $0.949-1.024  M_{\odot}$ are color-coded for each stellar mass and the values are indicated for  each group.  The  solid lines  correspond to  thick
envelope  sequences, while thinner  envelopes,
i.e. with $M_{\rm  H}/M_* < 2 \times 10^{-6}$, are  depicted with
different   lines,  with   increasing  $\log   g$  when   $M_{\rm  H}$
decreases.  The
figure   includes    cooling   curves   with   $1    M_{\odot}$   from
\citet{2001PASP..113..409F}. The spectroscopic  mass, determined using our evolutionary tracks, results in $\sim 0.974  M_{\odot}$ for  the sequences
with the thickest envelope, 4.3 \%  lower than the dynamical mass, in agreement with previous determinations of the spectrocopic mass \citep{2005MNRAS.362.1134B, 2012AJ....143...68H, 2017ApJ...848...11B, 2018MNRAS.tmp.1366J}. The  spectroscopic stellar mass  for each
hydrogen envelope mass, computed using the {\tt LPCODE} cooling tracks is listed in table \ref{masa-40}. Also listed, are stellar mass determinations from observations.


Also from Fig. \ref{SiriusB-1}, while the
cooling  track from \citet{2001PASP..113..409F} for  thin hydrogen  envelope (1.0-10) overlaps  with the  cooling sequences  of {\tt  LPCODE} with  the same mass, the track with thick hydrogen envelope (1.0-04, $10^{-4}$) overlaps with
the   {\tt    LPCODE}   tracks    with   stellar   mass    of   $0.973 M_{\odot}$.  We can easily explain the  difference in $\log g$  with the different total hydrogen mass in the models, since, for our models, the  thickest  hydrogen envelope  mass for  a $\sim 1  M_{\odot}$ white  dwarf is  $\sim 2\times  10^{-6} M_*$, two orders  of magnitude thinner than the value adopted by \citet{2001PASP..113..409F} (see section \ref{comparison}). We compute an additional sequence with $0.988 M_{\odot}$ and thick hydrogen envelope of $\sim 10^{-4} M_*$, labelled as 0998-thick in figure \ref{SiriusB-1}. 
We use the same technique described in section \ref{acretion-1} but, we turned off all hydrogen nuclear reactions, to keep its hydrogen content fixed. Note that  with an hydrogen envelope $\sim  100$ times more massive our model with  $0.998 M_{\odot}$ is  able to
nearly  reproduce  the spectroscopic surface  gravity  for  Sirius  B.  Although the hydrogen content is the dominant factor,  additional discrepancies  in the  surface  gravities between  the thick  envelope models can  be explained  with the difference  in the  helium content, being   arbitrarily  set   to   $10^{-2}M_*$  for   the  models   from \citet{2001PASP..113..409F},  and being  set by  stellar evolution  to
$10^{-3.1}M_*$ for the models computed with {\tt LPCODE}.

Note that, the uncertainties associated to the spectroscopic determinations
of the atmospheric parameters in the literature correspond to internal
errors and could  be as large as 1.2 \%  in effective temperature, and
0.038     dex     in      $\log     g$     \citep{2005MNRAS.362.1134B,
  2005ApJ...630L..69L}. In the case of Sirius B, \cite{2018MNRAS.tmp.1366J} computed the atmospheric parameters using different spectra for {\it HST} and found a dispersion for $\log g$ of 0.05 dex, leading to spectroscopic stellar masses between 0.874 and 0.962 $M_{\odot}$. With  this criteria, the uncertainties in the spectroscopic mass are three times larger than the ones considered by \cite{2017ApJ...840...70B}.  In addition, the  uncertainties presented by \citet{2017ApJ...840...70B} correspond to
uncorrelated internal  uncertainties of  the fitting, even  though the orbital parameters  and stellar masses  are correlated. By computing the uncertainties 
using a simple error propagation statistics, we obtain an uncertainty $\sim 48$\% larger for the mass of  Sirius B, implying    that   the   quoted    uncertainties could be underestimated.

\begin{figure}
\includegraphics[width=\columnwidth]{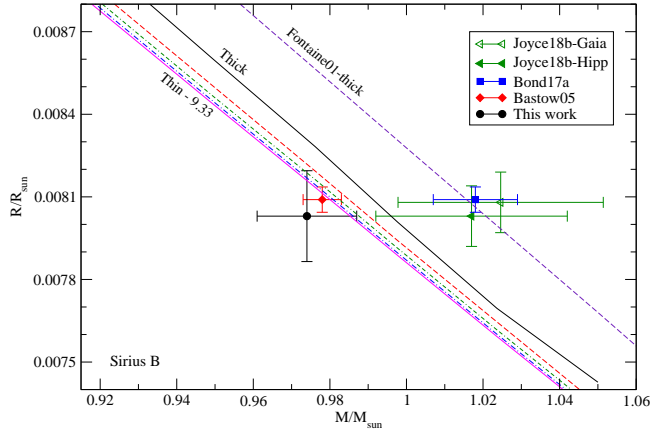}
\caption{Same  as  figure  \ref{40EriB-2}  but for  Sirius  B  and  an  effective  temperature of  $T_{\rm eff}=  25\, 369$ K. The points show the location of the spectroscopic mass obtain in this work combined with the radius from \citet{2018MNRAS.tmp.2291J} (circle), the dynamical    mass   from    \citet{2017ApJ...840...70B} (square) and the gravitational redshift mass from \citet{2018MNRAS.tmp.2291J} considering parallax from {\it Hipparcos} (full triangle) and {\it Gaia} DR2 (open triangle). 
\label{SiriusB-2}}
\end{figure}

In Figure \ref{SiriusB-2} we  compare the observational parameters for
Sirius  B   with  our   theoretical  models  using   the  mass--radius
relation. The different lines correspond to theoretical mass--radius relations for an effective temperature of $T_{\rm eff}= 25\, 369$ K. The solid black line corresponds to the sequences with the thickest hydrogen envelope allowed by single stellar evolution, computed with {\tt LPCODE}, while the solid magenta line correspond to the thinnest envelope, with $M_{\rm H} \sim 10^{-9.33} M_*$. We also show the theoretical mass--radius relation from \citet{2001PASP..113..409F} with hydrogen envelope mass $M_{\rm H} \sim 10^{-4} M_*$ as a dashed line. We include the gravitational redshift mass from \cite{2018MNRAS.tmp.2291J} obtained using parallaxes from {\it Hiparcos} (full triangle) and {\it Gaia} DR2 (open triangle), from \citet{2017ApJ...840...70B} (full square) and from \citet{2005MNRAS.362.1134B} (red diamond). With a black circle, we show the result obtained by considering the spectroscopic mass computed in this work combined with the radius from \cite{2018MNRAS.tmp.2291J}. 

As expected, the spectroscopic mass and the dynamical mass from \cite{2017ApJ...840...70B} do not agree within the uncertainties in 1 $\sigma$. However, the results from \cite{2018MNRAS.tmp.2291J} are compatible with our theoretical mass -- radius relation, within 1 $\sigma$ for a hydrogen  envelope with $M_H \sim 2\times 10^{-6} M_*$. Note that, the mass and radius for Sirius B from  \cite{2018MNRAS.tmp.2291J} are also in agreement with the ``thick'' envelope, with $M_H = 10^{-4} M_*$, sequences from \citep{2001PASP..113..409F}.

A thicker  envelope could perhaps be expected if Sirius B  had accretion
episodes   after    the   residual   nuclear   burning    has   turned
off. Considering that the  Sirius system  is a  visual binary  with a
period of $\sim 50$  yr \citep{2017ApJ...840...70B}, this scenario can
be disregarded. The  accretion rate of hydrogen  from the interstellar
medium is less than $10^{-17} M_{\odot}$/yr \citep{1993ApJS...84...73D,2015A&A...583A..86K}, too  low to  build a thick hydrogen envelope of $\sim 10^{-4} M_{*}$. In any case, as it was shown in section \ref{acretion-1}, the increase of the 
hydrogen content will trigger nuclear burning at the base of the envelope, 
reducing the hydrogen mass to $\sim 10^{-6}M_*$. 



\citet{2011PASA...28...58D}  determined  the structure parameters  for Sirius A using photometry  and spectroscopy combined with  parallax, to  be $R=1.7144 \pm  0.009 R_{\odot}$,  $T = 9845\pm 64$ K  and $L=24.74 \pm 0.70 L_{\odot}$. Considering the uncertainties in mass and metallicity, we estimate an age between $205  - 245$ Myr. With a cooling age of $115
\pm 6$ Myr,  the stellar mass of  the progenitor of Sirius  B is $5.11
^{+0.47}_{-0.28}M_{\odot}$, in  agreement with  the value  obtained by
\citet{2017ApJ...840...70B} and \citet{2005ApJ...630L..69L}.




\subsubsection{Binaries with non-DA white dwarf components}

In this section we briefly consider two non-DA white dwarfs in binary systems: Procyon B and Stein 2051 B. Procyon B is a DQZ white dwarf with an effective temperature of $7740\pm 50$ K  \citep{2002ApJ...568..324P} in a binary system with a slightly evolved subgiant of spectral type F5 IV-V. The Procyon system was analysed by \citet{2015ApJ...813..106B} using precise relative astrometry for {\it HST} observations combined with ground base observations and parallax. For Procyon B, the dynamical mass resulted in $M_* = 0.593 \pm 0.006 M_{\odot}$ and the radius, determined using flux and parallax measurements, was $0.01232\pm 0.00032 R_{\odot}$.

Stein 2051 B, is a DC white dwarf with $T_{\rm eff} = 7122\pm 181$ K, in a binary system with a main sequence companion of spectral type M4. The stellar mass was determined by \citet{2017Sci...356.1046S} using astrometric microlensing, being $M=0.75\pm 0.051 M_{\odot}$, while the radius of $R_* = 0.0114\pm 0.0004 R_{\odot}$ was determine using photometry and parallax measurements.

\begin{figure}
\includegraphics[width=\columnwidth]{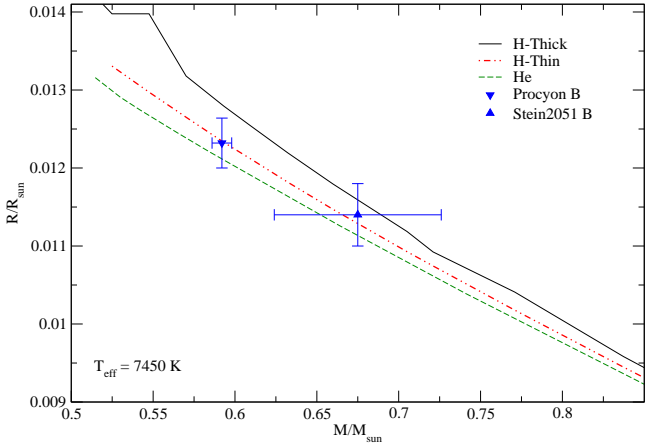}
\caption{Same as figure \ref{40EriB-2} but for the non-DA white dwarfs Procyon B and Stein 2051 B, where the observations are those from \citet{2015ApJ...813..106B} and \citet{2017Sci...356.1046S}, respectively. The curves are the mass--radius relation  for an effective temperature $T_{\rm eff} = 7450$ K, for thick (solid balck line) and thin ($M_H = 10^{-9}M_*$, red point-dashed line) and DB white dwarf models from  \citet{2009ApJ...704.1605A} (green dashed line). \label{dbs}}
\end{figure}

Figure \ref{dbs} shows the position of Procyon B and Stein 2051 B as compared to the theoretical mass--radius relations for $T_{\rm eff} = 7450$ K. The red point-dashed line corresponds to thin hydrogen envelope models, with $M_H = 10^{-9} M_*$ while the green dashed line corresponds to DB white dwarf models from \citet{2009ApJ...704.1605A}. We also included the mass-radius relation for thick envelope models computed with {\tt LPCODE}, those with the thickest hydrogen envelope allowed by stellar evolution. Considering the uncertainties reported by \citet{2015ApJ...813..106B}, Procyon B is in very good agreement with our theoretical models for thin H-envelope, as it was found by \citet{2015ApJ...813..106B, 2017ApJ...848...16B}, but also is in agreement with the theoretical mass--radius relation for DB white dwarfs. The results for Stein 2051 B are not that conclusive since the uncertainties are too large, but are still consistent with the theoretical models.

\subsection{Eclipsing binaries}

\begin{figure*}
\includegraphics[width=0.8\textwidth]{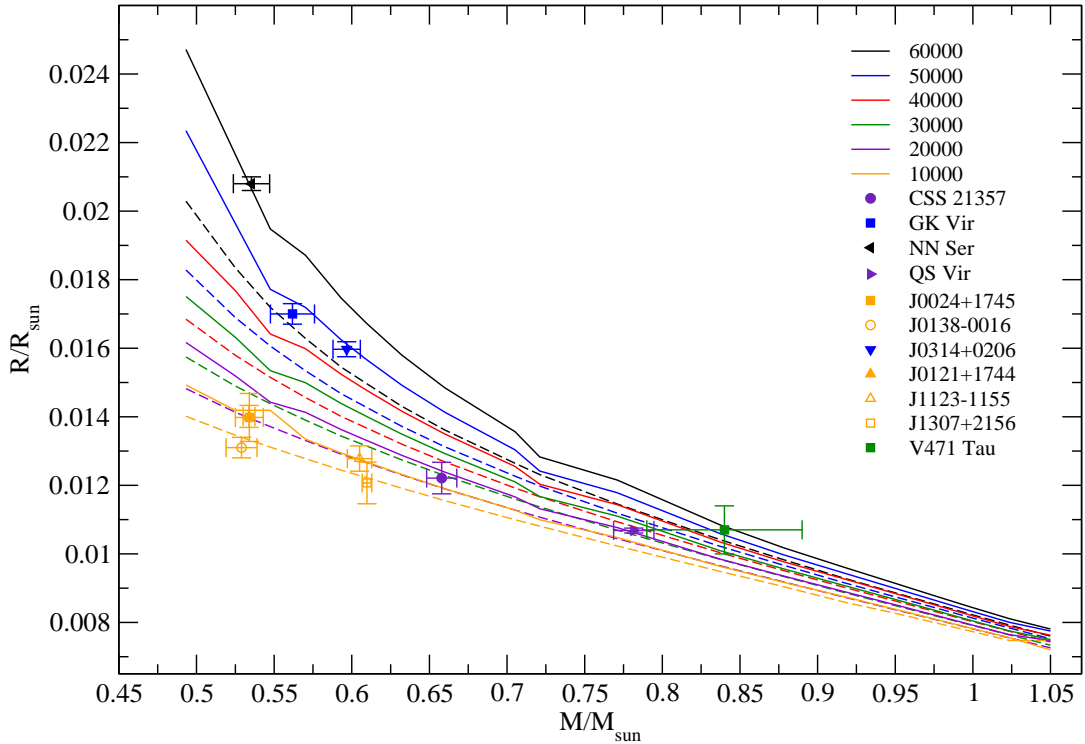}
\caption{The white dwarf Mass-Radius relation. Solid lines correspond to thick envelope models, while dashed lines correspond to 
sequences with $M_H/M_* = 10^{-10}$. From top to bottom, the solid lines are color coded, corresponding to effective temperatures of 
$60\, 000$ K (black), $50\, 000$ K (blue), $40\, 000$ K (red), $30\, 000$ K (green), $20\, 000$ K (violet), $10\, 000$ K (orange). 
The data point are the sample extracted from \citet{2017MNRAS.470.4473P} and indicated with different symbols that
are color coded as well. \label{parsons17}}
\end{figure*}

\citet{2017MNRAS.470.4473P}  presented mass and radius determinations for 16 white dwarfs in detached eclipsing binaries 
with low mass main sequence stars companions and combined them with 10 previous measurements to test the theoretical mass--radius relation. 
The mass and radius are estimated from the eclipses and radial velocity measurements, while the effective temperature of the white dwarf
component is determine using spectroscopy. 
We selected the objects with stellar masses larger than $\sim 0.5 M_{\odot}$, that are covered by our model grid. The selected sample is depicted in figure \ref{parsons17}, were we compare the observations extracted from \citet{2017MNRAS.470.4473P} to the theoretical mass--radius relation. The solid (dashed) lines correspond to the theoretical mass--radius relations for canonical ($M_H/M_{\odot} = 10^{-10}$) models for effective temperatures from $60\, 000$ K to $10\, 000$ K, from top to bottom, in steps of $10\, 000$ K (see figure for details). From this figure we see a very good agreement between models and observations, being also consistent in effective temperature. 

\begin{table*}
\caption{White dwarf stars in detached eclipsing binaries analysed in figure \ref{parsons17-fits}. For each object we list the effective temperature, stellar mass and radius (columns 2, 3 and 4) from \citet{2017MNRAS.470.4473P} and the hydrogen envelope mass determined in this work.  }  \centering
\begin{tabular}{ccccc}
\hline\hline 
obj ID  & $T_{\rm eff} (K)$  & $M (M_{\odot})$  & $R (R_{\odot})$ & $ M_H/M_*$   \\
\hline
CSS 21357  & $15909\pm  285$ & $0.6579\pm  0.0097$ & $0.01221\pm  0.00046$ & $5.8\times 10^{-5} - 4.6\times 10^{-10}$\\
GK Vir     & $50000\pm  673$ & $0.5618\pm  0.0142$ & $0.01700\pm  0.00030$ & $(1.67\pm 0.30)\times 10^{-4}$\\
NN Ser     & $63000\pm 3000$ & $0.5354\pm  0.0117$ & $0.02080\pm  0.00020$ & $(2.18\pm 0.23)\times 10^{-4}$\\
QS Vir     & $14220\pm  350$ & $0.7816\pm  0.0130$ & $0.01068\pm  0.00007$ & $1.8\times 10^{-5} - 4.6\times 10^{-10}$\\
J0024+1745 & $ 8272\pm 580$ & $0.5340\pm  0.0090$ & $0.01398\pm  0.00070$  & $2.2\times 10^{-4} - 2.6\times 10^{-10}$\\
J0138-0016 & $ 3570\pm 100$ & $0.5290\pm  0.0100$ & $0.01310\pm  0.00030$  & $2.3\times 10^{-4} - 5.8\times 10^{-10}$\\
J0314+0206 & $46783\pm 7706$ & $0.5967\pm  0.0088$ & $0.01597\pm  0.00022$ & $(1.25\pm 0.01)\times 10^{-4}$\\
J0121+1744 & $10644\pm 1721$ & $0.5338\pm  0.0038$ & $0.01401\pm  0.00032$ & $(2.21\pm 0.73)\times 10^{-4}$\\
J1123-1155 & $10210\pm   87$ & $0.6050\pm  0.0079$ & $0.01278\pm  0.00037$ & $(1.02\pm 0.10)\times 10^{-4}$\\
J1307+2156 & $ 8500\pm  500$ & $0.6098\pm  0.0031$ & $0.01207\pm  0.00061$ & $9.6\times 10^{-5} - 1.4\times 10^{-10}$\\
V471 Tau   & $34500\pm 1000$ & $0.8400\pm  0.0500$ & $0.01070\pm  0.00070$ & $9.8\times 10^{-6} - 4.8\times 10^{-10}$\\  
\hline\hline
\label{parson17-table}
\end{tabular}
\end{table*}
  
\begin{figure}
\includegraphics[width=0.45\textwidth]{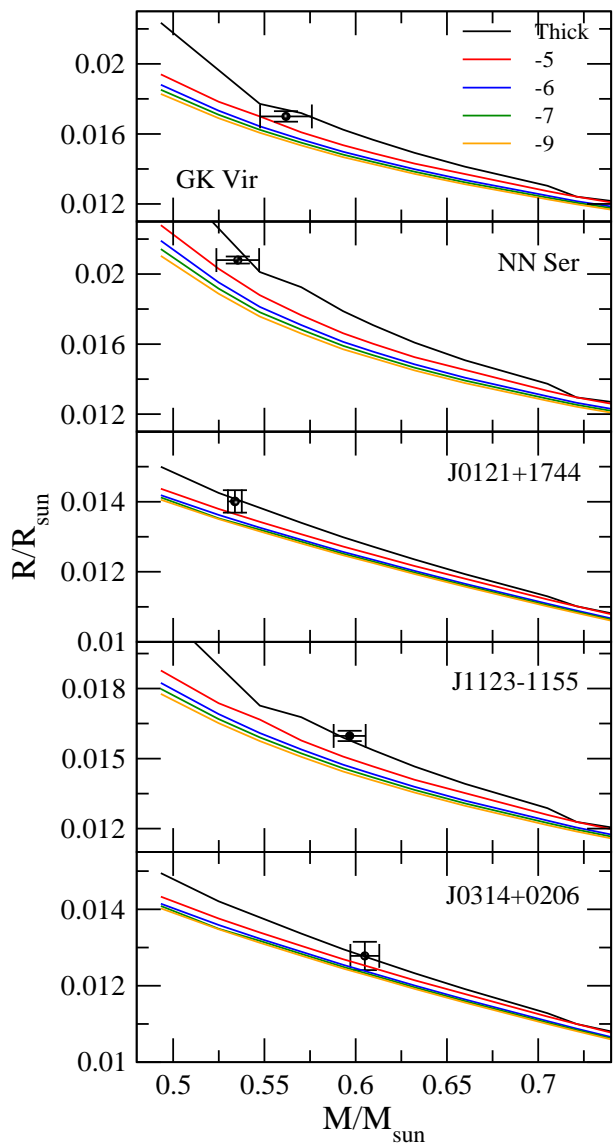}
\caption{Comparision with the observational values for mass and radius for five white dwarf in eclipsing binaries presented in \citet{2017MNRAS.470.4473P}, with the theoretical mass-radius relations for different masses of the hydrogen envelope. Each panel corresponds to a single object. The hydrogen envelope for each curve is color-coded (see inset in the figure). \label{parsons17-fits}}
\end{figure}

Next we proceed to measure the hydrogen content in the selected sample of eleven white dwarfs. The sample is listed in table \ref{parson17-table}, along with the effective temperature, stellar mass and radius extracted from \citet{2017MNRAS.470.4473P}. 
For each object we compare the observed mass and radius with our theoretical mass--radius relation, considering different thickness of the hydrogen envelope. From the selected sample, only five objects, GK Vir, NN Ser, J0138-0016, J0121+1744 and J1123-1155, show uncertainties small enough to measure the hydrogen envelope mass, within our model grid. The remaining objects are consistent with the theoretical mass--radius relation but the uncertainties are too large to constrain the mass of the hydrogen content.
The results for the five objects are depicted in figure \ref{parsons17-fits}, while the values for the hydrogen envelopes are listed in the last column of table \ref{parson17-table}. From figure \ref{parsons17-fits} it shows that all five objects have a canonical hydrogen envelope, i,e., the maximum amount of hydrogen as predicted by stellar evolution theory. This is expected given the mass range of the objects, for which the hydrogen envelope is intrinsically thicker. Also, note that the larger differences between the theoretical mass--radius relations for different hydrogen envelope mass occurs for low stellar masses and higher effective temperatures, as shown in figure \ref{tracks1}. 

\section{Conclusions}\label{end}

In this work we studied the mass--radius relation for white dwarf stars  and its dependence with the hydrogen envelope mass. In particular, how the extension of the hydrogen envelope affects the radius, and the surface gravity, which directly impacts the calculation of the stellar mass using atmospheric parameters, i.e., the spectroscopic mass.

We find that, comparing the sequences with the thickest envelope 
with those having the thinnest hydrogen envelope in our model grid, the reduction in the radius is around 8-12\%, 5-8\% and 1-2\% for stellar masses of 0.493 $M_{\odot}$, 0.609 $M_{\odot}$ and 0.998 $M_{\odot}$, respectively. As expected the differences are larger for models with lower stellar mass since, for these objects, the maximum hydrogen envelope left on top of a white dwarf star is thicker, according to single stellar evolution theory. 
The reduction of the stellar radius translates directly into an increase in $\log g$ for a fixed stellar mass, that can reach up to 0.11 dex, for low mass and high effective temperatures. Considering that the mean uncertainty in $\log g$ is 0.038 dex, then it is possible to measure the hydrogen envelope mass.

In addition, the maximum hydrogen mass allowed by stellar evolution theory is mass dependent, being
thinner than $M_H=10^{-4}M_*$, for white dwarf masses larger than $0.6 M_{\odot}$, and thicker for masses below $0.6 M_{\odot}$. Thus, considering a hydrogen envelope of $M_H=10^{-4}M_*$ for all stellar masses, can lead to overestimated or underestimated spectroscopic masses.

The hydrogen envelope mass is the dominant factor influencing the value of the radius. The central composition leads to less than 1\% difference in the radius, as shown by \citet{2017MNRAS.465.2849T}. For the helium content, the radius can be reduced by $\sim 1.1\%$ if the helium mass is reduced by more than a factor of 10.

We also use the mass--radius relation as a tool to measure the mass of the hydrogen envelope. We analyse a sample of white dwarf in astrometric and eclipsing binaries, for which it is possible to determine the mass and radius independently of the theoretical models. We consider four white dwarfs in astrometric binaries with very well determined orbital parameters and a sample of eleven white dwarf stars in detached eclipsing binary systems. Our main results are the following. 

\begin{itemize}

\item  For 40 Eridani B, we find a spectroscopic mass of $(0.573 \pm 0.0011) M_{\odot}$ and a hydrogen envelope mass of $M_H \sim 2.63\times 10^{-8} M_{\odot}$. This result is in agreement with previous determinations, pointing to a thin hydrogen envelope solution. The  cooling age for 40  Eridani B is $\sim  145$ Myr, with  a  total age  of $2.68\pm  0.43$ Gyr  and  a progenitor mass of $1.53\pm 0.11 M_{\odot}$.

\item For Sirius B, we find a spectrocopic stellar mass of $(0.974 \pm 0.013 M_{\odot}$ in agreement with previous determinations \citep{2005MNRAS.362.1134B}. In addition, the gravitational redshift mass from  \citet{2018MNRAS.tmp.2291J} as compared with our theoretical mass--radius relation, are in agreement within 1 $\sigma$, considering a thick hydrogen envelope of $M_H = 2.02 \times 10^{-6}$. The cooling age for Sirius B
is $115 \pm 6$ Myr, leading to a stellar mass of the progenitor of $5.11^{+0.47}_{-0.28}M_{\odot}$.
As compared to the dynamical mass, the spectrocopic value is 4.3\% lower than that obtained by \citet{2017ApJ...840...70B}, and not compatible within the uncertainties. We conclude that, either  the uncertainties in the dynamical mass are underestimated by at least $\sim$ 50\% or the difference is due to the fitting method and/or current atmospheric models.

\item Observations of both non-DA white dwarfs, Procyon B and Stein 2051 B, are consistent with a thin hydrogen envelope ($M_H \lesssim 10^{-9}$) as found by \citet{2017ApJ...848...16B}, but also with the pure He theoretical models from \citet{2009ApJ...704.1605A}.

\item For a sample of 11 white dwarfs in detached eclipsing binaries we found a good agreement between the theoretical mass--radius relation and the observations. For five objects, in the low mass range ($ \lesssim 0.6 M_{\odot}$), we measured the hydrogen mass and found thick hydrogen envelopes in all cases. For the remaining objects the uncertainties are too large to constrain the hydrogen envelope mass, but the observations are in agreement with the theoretical mass--radius relation.

\end{itemize}

In general the mass--radius relation computed using our models is in good agreement with the observations. For some objects we were able to constrain the hydrogen envelope mass given the lower uncertainties in the observed mass and radius. However, for most objects uncertainties are still too large. High mass white dwarf models show that these stars are born with hydrogen envelopes of $\sim 10^{-6} M_*$ or thinner. Thus the challenge of constrain the hydrogen mass is higher since the difference in radius is 1-2\% within the hydrogen mass range allowed by our model grid. 

Finally, we emphasize that the maximum hydrogen content left on top of a white dwarf is mass dependent, when the evolution of the white dwarf progenitor in computed consistently. In particular, the hydrogen envelope is thinner than the canonical value for stellar masses larger than $0.6 M_{\odot}$ and thicker for stellar masses below that value. Not taking into account this dependence can lead to a overestimation of the stellar mass when the determination is based on spectroscopy, i.e., using the atmospheric parameters $\log g$ and effective temperature.

\section*{Acknowledgements}
We thank our anonymous referee for the constructive
comments and suggestions.
ADR, SOK and GRL had financial support from CNPq and PRONEX-FAPERGS/CNPq
(Brazil). SRGJ acknowledges support from the Science and Technology Facilities Council (STFC, UK). AHC had financial support by AGENCIA through the Programa de Modernizaci\'on Tecnol\'ogica BID 1728/OC-AR, and by the PIP 112-200801-00940 grant from CONICET. 
Special thanks to Leandro Althaus for computing the model described in figure 5 and for the very useful comments on the manuscript.
This research has made use of NASA Astrophysics Data System. 




\begin{thebibliography}{}

\bibitem[Alexander \& Ferguson(1994)]{1994ApJ...437..879A} Alexander, D.~R., \& Ferguson, J.~W.\ 1994, \apj, 437, 879 
\bibitem[Althaus et al.(2003)]{2003A&A...404..593A} Althaus, L.~G., Serenelli, A.~M., C{\'o}rsico, A.~H., \& Montgomery, M.~H.\ 2003, \aap, 404, 593 
\bibitem[Althaus et al.(2005)]{2005A&A...435..631A} Althaus, L.~G., Serenelli, A.~M., Panei, J.~A., et al.\ 2005, \aap, 435, 631 
\bibitem[Althaus et al.(2009)]{2009ApJ...704.1605A} Althaus, L.~G., Panei, J.~A., Miller Bertolami, M.~M., et al.\ 2009, \apj, 704, 1605 
\bibitem[Althaus et al.(2010)]{2010ApJ...717..897A} Althaus, L.~G., C{\'o}rsico, A.~H., Bischoff-Kim, A., et al.\ 2010, \apj, 717, 897 
\bibitem[Althaus et al.(2013)]{2013A&A...557A..19A} Althaus, L.~G., Miller Bertolami, M.~M., \& C{\'o}rsico, A.~H.\ 2013, \aap, 557, A19 
\bibitem[Althaus et al.(2015)]{2015A&A...576A...9A} Althaus, L.~G., Camisassa, M.~E., Miller Bertolami, M.~M., C{\'o}rsico, A.~H., \& Garc{\'{\i}}a-Berro, E.\ 2015, \aap, 576, A9 
\bibitem[Barstow et al.(2005)]{2005MNRAS.362.1134B} Barstow, M.~A., Bond, H.~E., Holberg, J.~B., et al.\ 2005, \mnras, 362, 1134 
\bibitem[B{\'e}dard et al.(2017)]{2017ApJ...848...11B} B{\'e}dard, A., Bergeron, P., \& Fontaine, G.\ 2017, \apj, 848, 11 
\bibitem[Bergeron et al.(2001)]{2001ApJS..133..413B} Bergeron, P., Leggett, S.~K., \& Ruiz, M.~T.\ 2001, \apjs, 133, 413 
\bibitem[Bond et al.(2015)]{2015ApJ...813..106B} Bond, H.~E., Gilliland, R.~L., Schaefer, G.~H., et al.\ 2015, \apj, 813, 106 
\bibitem[Bond et al.(2017a)]{2017ApJ...840...70B} Bond, H.~E., Schaefer, G.~H., Gilliland, R.~L., et al.\ 2017a, \apj, 840, 70 
\bibitem[Bond et al.(2017b)]{2017ApJ...848...16B} Bond, H.~E., Bergeron, P., \& B{\'e}dard, A.\ 2017b, \apj, 848, 16 
\bibitem[Burgers(1969)]{1969fecg.book.....B} Burgers, J.~M.\ 1969, Flow Equations for Composite Gases, New York: Academic Press, 1969,  
\bibitem[Cassisi et al.(2007)]{2007ApJ...661.1094C} Cassisi, S., Potekhin, A.~Y., Pietrinferni, A., Catelan, M., \& Salaris, M.\ 2007, \apj, 661, 1094 
\bibitem[Castanheira \& Kepler(2009)]{2009MNRAS.396.1709C} Castanheira, B.~G., \& Kepler, S.~O.\ 2009, \mnras, 396, 1709 
\bibitem[Catal{\'a}n et al.(2008)]{2008MNRAS.387.1693C} Catal{\'a}n, S., Isern, J., Garc{\'{\i}}a-Berro, E., \& Ribas, I.\ 2008, \mnras, 387, 1693 
\bibitem[Cummings et al.(2016)]{2016ApJ...818...84C} Cummings, J.~D., Kalirai, J.~S., Tremblay, P.-E., \& Ramirez-Ruiz, E.\ 2016, \apj, 818, 84 
\bibitem[Davis et al.(2011)]{2011PASA...28...58D} Davis, J., Ireland, M.~J., North, J.~R., et al.\ 2011, \pasa, 28, 58 
\bibitem[De Ger{\'o}nimo et al.(2017)]{2017A&A...599A..21D} De Ger{\'o}nimo, F.~C., Althaus, L.~G., C{\'o}rsico, A.~H., Romero, A.~D., \& Kepler, S.~O.\ 2017, \aap, 599, A21 
\bibitem[De Ger{\'o}nimo et al.(2018)]{2018A&A...613A..46D} De Ger{\'o}nimo, F.~C., Althaus, L.~G., C{\'o}rsico, A.~H., Romero, A.~D., \& Kepler, S.~O.\ 2018, \aap, 613, A46 
\bibitem[Dupuis et al.(1993)]{1993ApJS...84...73D} Dupuis, J., Fontaine, G., Pelletier, C., \& Wesemael, F.\ 1993, \apjs, 84, 73 
\bibitem[El-Badry et al.(2018)]{2018ApJ...860L..17E} El-Badry, K., Rix, H.-W., \& Weisz, D.~R.\ 2018, \apjl, 860, L17 
\bibitem[Falcon et al.(2012)]{2012ApJ...757..116F} Falcon, R.~E., Winget, D.~E., Montgomery, M.~H., \& Williams, K.~A.\ 2012, \apj, 757, 116 
\bibitem[Fontaine et al.(2001)]{2001PASP..113..409F} Fontaine, G., Brassard, P., \& Bergeron, P.\ 2001, \pasp, 113, 409 
\bibitem[Fontaine \& Brassard(2008)]{2008PASP..120.1043F} Fontaine, G., \& Brassard, P.\ 2008, \pasp, 120, 1043 
\bibitem[Garcia-Berro et al.(1988)]{1988Natur.333..642G} Garcia-Berro, E., Hernanz, M., Isern, J., \& Mochkovitch, R.\ 1988, \nat, 333, 642 
\bibitem[Gatewood \& Gatewood(1978)]{1978ApJ...225..191G} Gatewood, G.~D., \& Gatewood, C.~V.\ 1978, \apj, 225, 191 
\bibitem[Haft et al.(1994)]{1994ApJ...425..222H} Haft, M., Raffelt, G., \& Weiss, A.\ 1994, \apj, 425, 222 
\bibitem[Heintz(1974)]{1974AJ.....79..819H} Heintz, W.~D.\ 1974, \aj, 79, 819 
\bibitem[Herwig et al.(1997)]{1997A&A...324L..81H} Herwig, F., Bloecker, T., Schoenberner, D., \& El Eid, M.\ 1997, \aap, 324, L81 
\bibitem[Holberg et al.(2012)]{2012AJ....143...68H} Holberg, J.~B., Oswalt, T.~D., \& Barstow, M.~A.\ 2012, \aj, 143, 68 
\bibitem[Horowitz et al.(2010)]{2010PhRvL.104w1101H} Horowitz, C.~J., Schneider, A.~S., \& Berry, D.~K.\ 2010, Physical Review Letters, 104, 231101 
\bibitem[Iben(1982)]{1982ApJ...260..821I} Iben, I., Jr.\ 1982, \apj, 260, 821 
\bibitem[Iben \& Renzini(1983)]{1983ARA&A..21..271I} Iben, I., Jr., \& Renzini, A.\ 1983, \araa, 21, 271 
\bibitem[Iben(1984)]{1984ApJ...277..333I} Iben, I., Jr.\ 1984, \apj, 277, 333 
\bibitem[Iben \& Tutukov(1984)]{1984ApJ...282..615I} Iben, I., Jr., \& Tutukov, A.~V.\ 1984, \apj, 282, 615 
\bibitem[Iglesias \& Rogers(1996)]{1996ApJ...464..943I} Iglesias, C.~A., \& Rogers, F.~J.\ 1996, \apj, 464, 943 
\bibitem[Istrate et al.(2014)]{2014A&A...571A..45I} Istrate, A.~G., Tauris, T.~M., \& Langer, N.\ 2014, \aap, 571, A45 
\bibitem[Itoh et al.(1996)]{1996ApJS..102..411I} Itoh, N., Hayashi, H., Nishikawa, A., \& Kohyama, Y.\ 1996, \apjs, 102, 411 
\bibitem[Joyce et al.(2018a)]{2018MNRAS.tmp.1366J} Joyce, S.~R.~G., Barstow, M.~A., Casewell, S.~L., et al.\ 2018a, \mnras,  479, 1612 
\bibitem[Joyce et al.(2018b)]{2018MNRAS.tmp.2291J} Joyce, S.~R.~G., Barstow, M.~A., Holberg, J.~B., et al.\ 2018b, \mnras,  481, 2361 
\bibitem[Koester et al.(1979)]{1979A&A....76..262K} Koester, D., Schulz, H., \& Weidemann, V.\ 1979, \aap, 76, 262 
\bibitem[Koester(2010)]{2010MmSAI..81..921K} Koester, D.\ 2010, \memsai, 81, 921 
\bibitem[Koester \& Kepler(2015)]{2015A&A...583A..86K} Koester, D., \& Kepler, S.~O.\ 2015, \aap, 583, A86 
\bibitem[Lauffer et al.(2018)]{2018MNRAS.480.1547L} Lauffer, G.~R., Romero, A.~D., \& Kepler, S.~O.\ 2018, \mnras, 480, 1547 
\bibitem[Liebert et al.(2005)]{2005ApJ...630L..69L} Liebert, J., Young, P.~A., Arnett, D., Holberg, J.~B., \& Williams, K.~A.\ 2005, \apjl, 630, L69 
\bibitem[Magni \& Mazzitelli(1979)]{1979A&A....72..134M} Magni, G., \& Mazzitelli, I.\ 1979, \aap, 72, 134 
\bibitem[Mason et al.(2017)]{2017AJ....154..200M} Mason, B.~D., Hartkopf, W.~I., \& Miles, K.~N.\ 2017, \aj, 154, 200 
\bibitem[Montgomery \& Winget(1999)]{1999ApJ...526..976M} Montgomery, M.~H., \& Winget, D.~E.\ 1999, \apj, 526, 976 
\bibitem[Parsons et al.(2010)]{2010MNRAS.402.2591P} Parsons, S.~G., Marsh, T.~R., Copperwheat, C.~M., et al.\ 2010, \mnras, 402, 2591 
\bibitem[Parsons et al.(2012a)]{2012MNRAS.426.1950P} Parsons, S.~G., G{\"a}nsicke, B.~T., Marsh, T.~R., et al.\ 2012a, \mnras, 426, 1950 
\bibitem[Parsons et al.(2012b)]{2012MNRAS.420.3281P} Parsons, S.~G., Marsh, T.~R., G{\"a}nsicke, B.~T., et al.\ 2012b, \mnras, 420, 3281 
\bibitem[Parsons et al.(2017)]{2017MNRAS.470.4473P} Parsons, S.~G., G{\"a}nsicke, B.~T., Marsh, T.~R., et al.\ 2017, \mnras, 470, 4473 
\bibitem[Paxton et al.(2011)]{2011ApJS..192....3P} Paxton, B., Bildsten, L., Dotter, A., et al.\ 2011, \apjs, 192, 3 
\bibitem[Paxton et al.(2013)]{2013ApJS..208....4P} Paxton, B., Cantiello, M., Arras, P., et al.\ 2013, \apjs, 208, 4 
\bibitem[Provencal et al.(1998)]{1998ApJ...494..759P} Provencal, J.~L., Shipman, H.~L., H{\o}g, E., \& Thejll, P.\ 1998, \apj, 494, 759 
\bibitem[Provencal et al.(2002)]{2002ApJ...568..324P} Provencal, J.~L., Shipman, H.~L., Koester, D., Wesemael, F., \& Bergeron, P.\ 2002, \apj, 568, 324 
\bibitem[Rebassa-Mansergas et al.(2007)]{2007MNRAS.382.1377R} Rebassa-Mansergas, A., G{\"a}nsicke, B.~T., Rodr{\'{\i}}guez-Gil, P., Schreiber, M.~R., \& Koester, D.\ 2007, \mnras, 382, 1377 
\bibitem[Renedo et al.(2010)]{2010ApJ...717..183R} Renedo, I., Althaus, L.~G., Miller Bertolami, M.~M., et al.\ 2010, \apj, 717, 183 
\bibitem[Romero et al.(2012)]{2012MNRAS.420.1462R}Romero, A.~D., C{\'o}rsico, A.~H., Althaus, L.~G., et al.\ 2012, \mnras, 420, 1462 
\bibitem[Romero et al.(2013)]{2013ApJ...779...58R} Romero, A.~D., Kepler, S.~O., C{\'o}rsico, A.~H., Althaus, L.~G., \& Fraga, L.\ 2013, \apj, 779, 58
\bibitem[Romero et al.(2015)]{2015MNRAS.450.3708R} Romero, A.~D., Campos, F., \& Kepler, S.~O.\ 2015, \mnras, 450, 3708 
\bibitem[Romero et al.(2017)]{2017ApJ...851...60R} Romero, A.~D., C{\'o}rsico, A.~H., Castanheira, B.~G., et al.\ 2017, \apj, 851, 60 
\bibitem[Sahu et al.(2017)]{2017Sci...356.1046S} Sahu, K.~C., Anderson, J., Casertano, S., et al.\ 2017, Science, 356, 1046 
\bibitem[Salaris et al.(1997)]{1997ApJ...486..413S} Salaris, M., Dom{\'{\i}}nguez, I., Garc{\'{\i}}a-Berro, E., et al.\ 1997, \apj, 486, 413 
\bibitem[Salaris et al.(2000)]{2000ApJ...544.1036S} Salaris, M., Garc{\'{\i}}a-Berro, E., Hernanz, M., Isern, J., \& Saumon, D.\ 2000, \apj, 544, 1036 
\bibitem[Salaris et al.(2010)]{2010ApJ...716.1241S} Salaris, M., Cassisi, S., Pietrinferni, A., Kowalski, P.~M., \& Isern, J.\ 2010, \apj, 716, 1241 
\bibitem[Schmidt(1996)]{1996A&A...311..852S} Schmidt, H.\ 1996, \aap, 311, 852 
\bibitem[Schr{\"o}der \& Cuntz(2005)]{2005ApJ...630L..73S} Schr{\"o}der, K.-P., \& Cuntz, M.\ 2005, \apjl, 630, L73 
\bibitem[Segretain et al.(1994)]{1994ApJ...434..641S} Segretain, L., Chabrier, G., Hernanz, M., et al.\ 1994, \apj, 434, 641 
\bibitem[Shipman et al.(1997)]{1997ApJ...488L..43S} Shipman, H.~L., Provencal, J.~L., H{\o}g, E., \& Thejll, P.\ 1997, \apjl, 488, L43
\bibitem[Tassoul et al.(1990)]{1990ApJS...72..335T} Tassoul, M., Fontaine, G., \& Winget, D.~E.\ 1990, \apjs, 72, 335 
\bibitem[Tremblay et al.(2017)]{2017MNRAS.465.2849T} Tremblay, P.-E., Gentile-Fusillo, N., Raddi, R., et al.\ 2017, \mnras, 465, 2849 
\bibitem[Tremblay et al.(2019)]{2019MNRAS.482.5222T} Tremblay, P.-E., Cukanovaite, E., Gentile Fusillo, N.~P., Cunningham, T., \& Hollands, M.~A.\ 2019, \mnras, 482, 5222 
\bibitem[van Horn(1968)]{1968ApJ...151..227V} van Horn, H.~M.\ 1968, \apj, 151, 227 
\bibitem[van Leeuwen(2007)]{2007A&A...474..653V} van Leeuwen, F.\ 2007, \aap, 474, 653 
\bibitem[Vassiliadis \& Wood(1993)]{1993ApJ...413..641V} Vassiliadis, E., \& Wood, P.~R.\ 1993, \apj, 413, 641 
\bibitem[Vauclair et al.(1997)]{1997A&A...325.1055V} Vauclair, G., Schmidt, H., Koester, D., \& Allard, N.\ 1997, \aap, 325, 1055 
\bibitem[Winget et al.(1987)]{1987ApJ...315L..77W} Winget, D.~E., Hansen, C.~J., Liebert, J., et al.\ 1987, \apjl, 315, L77 
\bibitem[Winget et al.(2009)]{2009ApJ...693L...6W} Winget, D.~E., Kepler, S.~O., Campos, F., et al.\ 2009, \apjl, 693, L6 
\bibitem[Wood(1995)]{1995LNP...443...41W} Wood, M.~A.\ 1995, White Dwarfs, 443, 41 




\end{thebibliography}







\bsp	
\label{lastpage}
\end{document}